\documentclass[iop]{emulateapj}
\usepackage{hyperref}
\usepackage{amsmath}
\usepackage{algorithm}
\usepackage{amsfonts}
\usepackage{mathrsfs}
\usepackage{bm}
\usepackage{rotating}
\usepackage{color}
\usepackage{graphicx}
\usepackage{subfigure}

\def\cha{\textit{Chandra}}
\def\XMM{{XMM-{\it Newton}}}
\def\NuSTAR{{\it NuSTAR}}
\def \XSPEC {{\tt XSPEC}}
\def \pexrav {{\tt pexrav}}
\def \MYTorus {{\tt MYTorus}}
\def \borus {{\tt borus02}}
\def \bntorus {{\tt BNtorus}}

\begin{document}
\title{Compton-thick AGN in the \NuSTAR\ era II: A deep \NuSTAR\ and \XMM\ view of the candidate Compton thick AGN in NGC 1358}
\author{X. Zhao\altaffilmark{1}, S. Marchesi\altaffilmark{1}, M. Ajello\altaffilmark{1}, L. Marcotulli\altaffilmark{1}, G. Cusumano\altaffilmark{2}, V. La Parola\altaffilmark{2}, C. Vignali\altaffilmark{3,4}}

\altaffiltext{1}{Department of Physics \& Astronomy, Clemson University, Clemson, SC 29634, USA}
\altaffiltext{2}{INAF - Istituto di Astrofisica Spaziale e Fisica Cosmica, Via U. La Malfa 153, I-90146 Palermo, Italy}
\altaffiltext{3}{INAF--Osservatorio Astronomico di Bologna, Via Piero Gobetti, 93/3, 40129, Bologna, Italy}
\altaffiltext{4}{Dipartimento di Fisica e Astronomia, Alma Mater Studiorum, Universit\`a di Bologna, Via Piero Gobetti, 93/2, 40129, Bologna, Italy}

\begin{abstract}
    We present the combined \NuSTAR\ and \XMM\ 0.6--79\,keV spectral analysis of a Seyfert 2 galaxy, NGC 1358, which we selected as a candidate Compton thick (CT-) active galactic nucleus (AGN) on the basis of previous \textit{Swift}/BAT and \cha\ studies. According to our analysis, NGC 1358 is confirmed to be a CT-AGN  using physical motivated models, at $>$3\,$\sigma$ confidence level. Our best-fit shows that the column density along the ``line-of-sight'' of the obscuring material surrounding the accreting super-massive black hole is N$\rm _H$ = [1.96--2.80] $\times$ 10$^{24}$\,cm$^{-2}$. The high-quality data from \NuSTAR\ gives the best constraints on the spectral shape above $\sim$10\,keV to date on NGC 1358. Moreover, by combining \NuSTAR\ and \XMM\ data, we find that the obscuring torus has a low covering factor ($f_c$ $<$0.17), and the obscuring material is distributed in clumps, rather than uniformly. We also derive an estimate of NGC 1358's Eddington ratio, finding it to be $\lambda_{\rm Edd}$ $\sim$$4.7_{-0.3}^{+0.3}$ $\times$ 10$^{-2}$, which is in acceptable agreement with previous measurements. Finally, we find no evidence of short-term variability, over a $\sim$100\,ks time-span, in terms of both ``line-of-sight'' column density and flux. 
\end{abstract}

\keywords{galaxies: active -- galaxies: nuclei -- galaxies: individual (NGC 1358) -- X-rays: galaxies}

%
%
\section{Introduction}\label{sec:intro}
The Cosmic X-ray Background (CXB; i.e., the diffuse X-ray emission observed between 0.5\,keV and 300\,keV) is thought to be mainly produced by obscured and unobscured active galactic nuclei  \citep[AGN; e.g.,][]{Alexander03,Gandhi03,gilli07,Treister09}. Compton-thick (CT-) AGNs (with absorbing column density N$\rm _H\ge\sigma_T^{-1}$ $\sim$10$^{24}$\,cm$^{-2}$, where $\sigma_T$ is the Thomson cross section) are supposed to contribute up to $\sim$10\% of the CXB intensity at its spectral peak \citep[$\sim$ 30\,keV,][]{Ajello08} and are expected to be numerous \citep[up to 50\% of the overall population of Seyfert 2 galaxies; see, e.g.,][]{risaliti1999}. However, as of today CT-AGNs have never been detected in large numbers, their observed fraction in the local Universe being $\sim$ 5--10\% \citep[see, e.g.,][]{Burlon11,Ricci15}, significantly below the predictions of different CXB models \citep[$\sim$20\%--30\%, see][and references therein]{Ueda14}. Nevertheless, it has been suggested that the small observed fraction of heavily obscured AGN observed can be caused by the bias in detecting CT-AGN in  X-rays, even sampling the energy range above 10\,keV \citep[see, e.g.][]{Burlon11}. Efforts to correct for this observational bias have recovered a fraction of $\sim$20\,\%  of CT-AGN, under some assumptions \citep[see, e.g.,][]{Burlon11,BNtorus, Ricci15}.

In Compton-thick AGN, the spectrum is significantly suppressed at energies $\le$10\,keV \citep{gilli07,koss2016} and the overall emission is dominated by the Compton hump at $\sim$30--50\,keV. Consequently, CT-AGNs at redshifts $z$ $>$1 can be studied using one of the several facilities sampling the $\sim$0.5--10\,keV energy range, such as \XMM, \cha, \textit{Swift}-XRT and \textit{Suzaku} \citep[see, e.g.,][]{Georgantopoulos2013,Buchner2015,Lanzuisi2105}, since the Compton hump of high-$z$ sources is redshifted in the energy range covered by these instruments. For sources in the local Universe ($z$ $<$0.1), however, the proper characterization of heavily obscured AGN requires an X-ray telescope sensitive above 10\,keV. Thanks to the launch of \textit{Nuclear Spectroscopic Telescope Array} \citep[hereafter, \NuSTAR,][]{harrison}, which  provides a two orders of magnitude better sensitivity than previous telescopes at these energies \citep[e.g., \textit{INTEGRAL} and \textit{Swift}/BAT;][]{Winkler2003,Barthelmy2005}, we can characterize the physical properties of heavily obscured AGN with unprecedented accuracy \citep[see, e.g.,][]{Balokovic14,puccetti14,Annuar15,Stefano2017,Ursini18}. 
However, since a typical highly obscured AGN spectrum barely depends on the column density at $>$10\,keV but varies considerably at $<$10\,keV \citep[see, e.g.,][]{gilli07}, it is difficult to constrain the column density with \NuSTAR\ alone. Consequently, \XMM, as the best instrument in terms of effective area in 0.3--10\,keV ($\sim$10 times larger than \textit{Swift}-XRT and $\sim$ 2 times larger than \cha), is the ideal instrument to complement \NuSTAR\ strength in characterizing heavily obscured AGNs.

Indeed, the study of single targets using \NuSTAR\ or combining \NuSTAR\ and other lower-energy X-ray observatories (e.g., \XMM\ and \cha) has already been shown to be strategical to characterize heavily obscured AGN, and understand their physical properties. For example, NGC 1448 was observed and identified as a CT-AGN in X-rays for the first time using \NuSTAR\ and \cha\ \citep{Annuar17}. The source was too faint (intrinsic 2--10\,keV luminosity $L_{\rm int, 2-10
~keV}$ = 3.5--7.6 $\times$ 10$^{40}$ erg s$^{-1}$) to be identified by \textit{Swift}/BAT, even using its deepest 104 month maps, and was only detected in one out of five \textit{Swift}-XRT observations. Another example is the analysis of NGC 1068 reported in \citet{Bauer15}. In this work, the authors used \NuSTAR\ to characterize with unprecedented quality this largely studied CT-AGN, putting much stronger constraints on the high-energy spectral shape of NGC 1068.

The obscuration observed in AGN across the electromagnetic spectrum, from the X-ray, to the optical and infrared, is usually explained with a pc-scale, torus-like structure of dust and gas \citep[see, e.g.,][]{Natureastro2017}. Consequently, in the past two decades several tori models, based on Monte Carlo simulations, have been developed to characterize CT-AGN X-ray spectra \citep{Matt1994,Shinya09,MYTorus2009,BNtorus,Liu14,Furui16,Borus}. All these models assume a continuous distribution of the obscuring material, but with different assumption on the geometry of the torus. In particular, in the models proposed by \citet{Shinya09}, \citet{BNtorus} and \citet{Borus}, the half opening angle of torus, i.e., the torus covering factor, is a free parameter, thus allowing to put constraints on the typical tori geometry. Given the intrinsic complexity of these models, and the multiple free parameters involved, using them in full capacity requires high-quality X-ray spectra, with excellent statistics on a wide energy range, i.e., between 2 and 100\,keV: as of today, such requirements can be satisfied only by a joint \NuSTAR\ and \XMM\ observation.

In this work we present the results of a deep, 50\,ks joint \NuSTAR\ and \XMM\ observation of NGC 1358, a nearby Seyfert 2 galaxy and a CT-AGN candidate. The paper is organized as follows: in Section \ref{sec:Observ} we present the selection technique that brought us to classify NGC 1358 as a new candidate CT-AGN, and we report the \NuSTAR\ and \XMM\ data reduction and spectral extraction process. In Section \ref{sec:spectral}, we describe the different models, both phenomenologicals and physicals, which have been used to fit the spectra, and the results of the spectral analysis.
In Section \ref{discussion} we compare our results with previous ones, derive the source Eddington ratio and discuss the constraints on the geometry and clumpiness of the obscuring materials.
All reported errors are at 90\% confidence level, if not otherwise stated. Standard cosmological constants are adopted as follows: $<H_0>$ = 70 km s$^{-1}$ Mpc$^{-1}$, $<q_0>$ = 0.0 and $<\Lambda>$ = 0.73.
%
%
\section{Observation and Data Analysis}\label{sec:Observ}
NGC 1358 \citep[z $\sim$0.013436,][]{Theureau1998}, is a Seyfert 2 galaxy detected in the 100-month BAT catalog (with a 7.8\,$\sigma$ significance; Segreto et al. 2018 in prep.), a catalog of $\sim$1000 AGN detected by \textit{Swift}-BAT in the 15--150\,keV band. 

In \citet{marchesi2017APJ}, we describe a technique developed to select highly obscured AGN candidates from the BAT sample, using the following criteria:
\begin{enumerate}
\item Lack of a 0.5--2.4\,keV, \textit{ROSAT}/RASS \citep{Boller16} counterpart. For objects located outside the Galactic plane (i.e., having Galactic latitude $\lvert$$b$$\rvert$$>$10 $\deg$), the lack of \textit{ROSAT} counterparts already implies a minimum AGN column density log(N$\rm _H$) $\sim$23 \citep[see, e.g., Figure 2 in][]{koss2016}.
\item Seyfert 2 galaxy optical classification, i.e., the source must have an optical spectrum without broad (FWHM $\geq$2000\,km s$^{-1}$) emission lines. It has been shown \citep[see, e.g.,][and references therein]{marchesi2016} that Seyfert 2 galaxies are more likely to be obscured than Seyfert 1 ones. Furthermore, there are no known Seyfert 1 galaxy that are Compton thick\footnote{there are sources which are Compton-thick but with ambiguous activity classification, e.g. NGC 424 (a.k.a. Tololo0109-383)} \citep[see, e.g.,][]{Ricci15}.
\item Low redshift ($z$ $<$0.04). Due to selection effects, the vast majority of BAT-selected CT-AGNs are detected in the nearby Universe: for example, 47 out of 55 CT-AGNs reported in \citet{Ricci15} are located at $z$ $<$0.04.
\end{enumerate}

Following these criteria, we obtained a snapshot (10\,ks) \cha\ observation for a sample of seven sources, and we performed a first measurement of their fundamental spectral parameters, particularly the power law photon index, $\Gamma$, and the column density, N$\rm _H$. 
NGC 1358 was found to be the most obscured object in our sample, having ``line-of-sight" column density N$\rm _H$ = 1.05$_{-0.36}^{+0.42}$ $\times$ 10$^{24}$ cm$^{-2}$, thus making it a candidate CT-AGN, although only at a 1\,$\sigma$ confidence level, due to the low-quality of the \cha\ spectrum. 

To further investigate this new candidate CT-AGN we proposed for a joint deep \NuSTAR\ (50\,ks) and \XMM\ (48\,ks) follow-up observation, which was accepted in \NuSTAR\ Cycle 3 (proposal ID 3258, PI: Marchesi). We report a summary of the two observations in Table \ref{tabel:1}.

\begin{table*}
\center
\caption{Summary of \NuSTAR\ and \XMM\ observation.}
\label{tabel:1}
\vspace{.1cm}
  \begin{tabular}{cccccc}
       \hline
       \hline
    Instrument&Sequence&Start Time&End Time&Exposure Time&Count Rate\tablenotemark{a}\\ 
    &ObsID&(UTC)& (UTC)&(ks)&$10^{-2}$counts s$^{-1}$\\
    \hline
    \NuSTAR&60301026002&2017-08-01T03:41:09&2017-08-02T06:36:09&50&2.32$\pm$0.07  2.28$\pm$0.07\\
    \XMM&0795680101&2017-08-01T17:05:27&2017-08-02T06:03:10&48&0.98$\pm$0.05 0.91$\pm$0.05 3.68$\pm$0.15\\
       \hline
\end{tabular}
\par
\vspace{.2cm}
\tablenotemark{a}{The reported \NuSTAR\ net count rates are those of the FPMA and FPMB modules between 3--79\,keV, respectively. The reported \XMM\ net count rates are those the MOS1, MOS2 and pn modules in 0.6--10\,keV, respectively.}
\end{table*}

\subsection{\NuSTAR\ Observation}
NGC 1358 was observed by \NuSTAR\ on 2017 August 1 (ObsID 60301026002): the net exposure time is 50\,ks. The observation actually took place in a 96.9\,ks time-span and was divided in 16 ($\sim$3\,ks) intervals. The non-exposed time between each interval is when the target is occulted by the Earth.

The \NuSTAR\ data are derived from both focal plane modules, FPMA and FPMB. The raw files are calibrated, cleaned and screened using the \NuSTAR\ \textit{nupipeline} script version 0.4.5. The \NuSTAR\ calibration database (CALDB) used in this work is the version 20161021. The ARF, RMF and light-curve files are obtained using the \textit{nuproducts} script. 

For both modules, the source spectrum is extracted from a 25$^{\prime\prime}$ circular region, corresponding to $\approx$40\% of the encircled energy fraction at 10 keV, centered on the source optical position. We then extract a background spectrum for each module, choosing a 30$^{\prime\prime}$ circular region located nearby the outer edges of the field of view, to avoid contamination from NGC 1358. We group the \NuSTAR\ spectra with a minimum of 15 counts per bin with $grppha$ task. The signal of both modules is $>$3\,$\sigma$ in 3--79\,keV band.

\subsection{\XMM\ Observation} 
The \XMM\ observation was taken quasi-simultaneously to the \NuSTAR\ one starting $\sim$12 hours after the \NuSTAR\ one, but ending at the same time (due to the gaps between observing intervals in \NuSTAR). \XMM\ data have been reduced using the Science Analysis System \citep[SAS;][]{SAS} version 16.1.0. 13\,ks of \XMM\ modules MOS1 and MOS2 and 30\,ks of pn observations were affected by a strong background flare, therefore we decided to exclude that part of observation from our analysis. Consequently, the total net \XMM\ exposure time of our observation is 101\,ks. The source spectra are extracted from a 15$^{\prime\prime}$, corresponding to $\approx$ 70\% of the encircled energy fraction at 1.5\,keV, circular region, while the background spectra are from a 80$^{\prime\prime}$ circle located nearby the source. We visually inspected the \XMM\ image to avoid contamination to the background from sources nearby NGC 1358. All three modules, MOS1, MOS2 and pn are jointly used in the spectral modeling, and their normalization are tied together assuming their cross-calibration uncertainties are marginal.

%
%
\section{Spectral Modeling Results}\label{sec:spectral}
We use \XSPEC\ \citep{Arnaud1996} v12.9.1 to fit the spectrum and rely on the $\chi^2$ statistic for the optimization of the spectral fit. The photoelectric cross section for all absorption components used here are derived from \cite{Verner1996}, adopting an element abundance from \citet{Anders1989}. The Galactic absorption column density is N$\rm _H^{Gal}$ = $3.83\times 10^{20}$\,cm$^{-2}$ \citep{Kalberla05}. The metal abundance is fixed to Solar.

Following a standard approach in analyzing heavily obscured AGN, we first fit our data using different phenomenological models, particularly the \pexrav\ one \citep{pexrav}. We then move to more accurate self-consistent models, based on Monte Carlo simulations, which are specifically developed to treat the spectra of heavily obscured AGN: the physical models we use in this work are \MYTorus\ \citep{MYTorus2009} and \borus\ \citep{Borus}. We report the results of our analysis in the following sections.

\subsection{Phenomenological Models}\label{pheno}

\subsubsection{Absorbed power law}\label{sec:model_abs}
We initially fit our data with a simple phenomenological model, comprising a power law (\textit{zpowerlw} in \XSPEC) absorbed by intervening gas modeled with (\textit{zphabs}). We also add a Gaussian (\textit{zgauss}) to model the Fe K$\alpha$ fluorescent emission line (E$_{\rm K\alpha}$ = 6.4\,keV); we assume the line to be narrow, fixing the line width $\sigma$ to 50\,eV, since there is no statistical improvement in fits if the parameter is left free to vary. 
We also add a second, unabsorbed power law, to model the fractional AGN emission, which is not intercepted by the torus on the ``line-of-sight'', and/or the scattering emission that is deflected, rather than absorbed by the obscuring material. Here, and elsewhere in the paper, the photon index of the scattered component is tied to  the one of the main power law. The scattered component is usually less than 5--10\% of the main one \citep[see, e.g.,][]{Marchesi2018}. We denote this fraction as $f_s$, and we model it with a constant ($constant_2$).
Furthermore, we add to the fit a thermal component, namely \textit{mekal} \citep{mekal}, to model the soft excess observed below 1\,keV, and potentially due to either star-formation processes and/or thermal emission from a hot interstellar medium. The temperature and the relative metal abundance in \textit{mekal} are both left free to vary.

The first model (hereafter, ``Model A''), in \XSPEC\ nomenclature, is therefore:
\begin{equation}
\label{eq:powerlaw}
\begin{aligned}
Model A=&constant_1*phabs*(zphabs*zpowerlw+\\
&zgauss+constant_2*zpowerlw+mekal)
\end{aligned}
\end{equation}
where $constant_1$ represents the cross calibration between different instruments, noted as $C_{NuS/XMM}$. In our fits, the cross-calibration between different modules of the same instrument is fixed to 1. \textit{phabs} is applied here to model the Galactic absorption. 

We report in Table \ref{Table:best-fit} the best-fit results for the simple phenomenological model applied to the joint \NuSTAR--\XMM\ spectrum. The best-fit photon index is $\Gamma$ = 1.14$_{-0.12}^{+0.13}$; the column density is N$_{\rm H}$ = 0.95$_{-0.11}^{+0.11}$ $\times$ $10^{24}$\,cm$^{-2}$.
While the best-fit reduced $\chi^2$ of model A is statistically acceptable, being $\chi^2_\nu$ = $\chi^2$/degree of freedom (d.o.f. hereafter) = 256/240 = 1.07, standard absorption components in \XSPEC, such as \textit{zphabs}, fail to characterize the spectral complexity of heavily obscured AGN like NGC 1358 properly. Therefore, a more physical model needs to be applied.

\subsubsection{Including a reflection component}\label{sec:pexrav}
Obscured AGNs X-ray spectra have historically been modeled using the \pexrav\ model \citep{pexrav}. \pexrav\ is used to model an exponentially cut-off power law spectrum reflected from neutral slab. We first test the model with a pure reflector by setting the reflection scaling factor in \pexrav\ to be $R$ = -1: this models a heavily obscured (N$_{\rm H}$ $>10^{25}$\,cm$^{-2}$) source whose spectrum is dominated by the reflection from the ``back-side'' of the torus.
The fit shows that photon index is $\Gamma$ = $1.30^{+0.05}_{-0.05}$ and $\chi^2$/d.o.f = 349/242. Such a large reduced $\chi^2$ suggests that a pure reflector is not enough to describe the spectrum. Therefore, we follow the method described in, e.g.,  \citet[][]{Ricci11} by using the complete \pexrav\ model, which includes an intrinsic cut-off power law by setting the reflection scaling factor $R$ to be greater than 0.

The model in \XSPEC\ is described as follows:
\begin{equation}\label{eq:pexrav}
\begin{aligned}
Model B=&constant_1*phabs*(zphabs*pexrav+\\
&zgauss+constant_2*zpowerlw+mekal)
\end{aligned}
\end{equation}

The components are those described previously in Section \ref{sec:model_abs}, except for the main power law, which is replaced by \pexrav. The inclination angle $i$, i.e., the angle between the axis of the AGN (normal to the disk) and the observer ``line-of-sight'', which is a free parameter in \pexrav, is fixed at $i$ = 60$^{\circ}$ (i.e. cos $i$ = 0.5): we find no significant change in the best-fit statistic and in the other parameters when allowing $i$ to vary. The cut-off energy of \pexrav\ is fixed at 500\,keV, to be consistent with the \MYTorus\ model, which we will extensively discuss in the following section. 

When leaving the reflection scaling factor $R$ in \pexrav\ free to vary, we obtain a  best-fit value of R $>$4 \cite[such a large reflection scaling factor is also found by][in heavily obscured AGN]{Ricci11}, although we are not able to put any constraint on the parameter 90\% confidence uncertainties. Such a result would point towards a ``reflection-dominated'' scenario, where most of the observed emission comes from the reflected component, while the direct emission from the accreting SMBH is absorbed by the heavily obscuring material along the ``line of sight''. A larger scaling factor can also be interpreted as the geometry of reflected material is more like a torus rather than a disk \footnote{as a larger value of the scaling factor represents a larger amount of reflected material, thus the reflected material is more torus-like rather than disk-like geometrically.}. Since $R$ is not constrained when left free to vary, we decided to complete our spectral analysis fixing the reflection scaling parameter to $R$ = 4. Here we are modeling a process that the intrinsic emission together with the reflection from the ``back-side'' are obscured by the same circumnuclear material.

\begin{figure}[htpb]
\centering
\includegraphics[width=.5\textwidth]{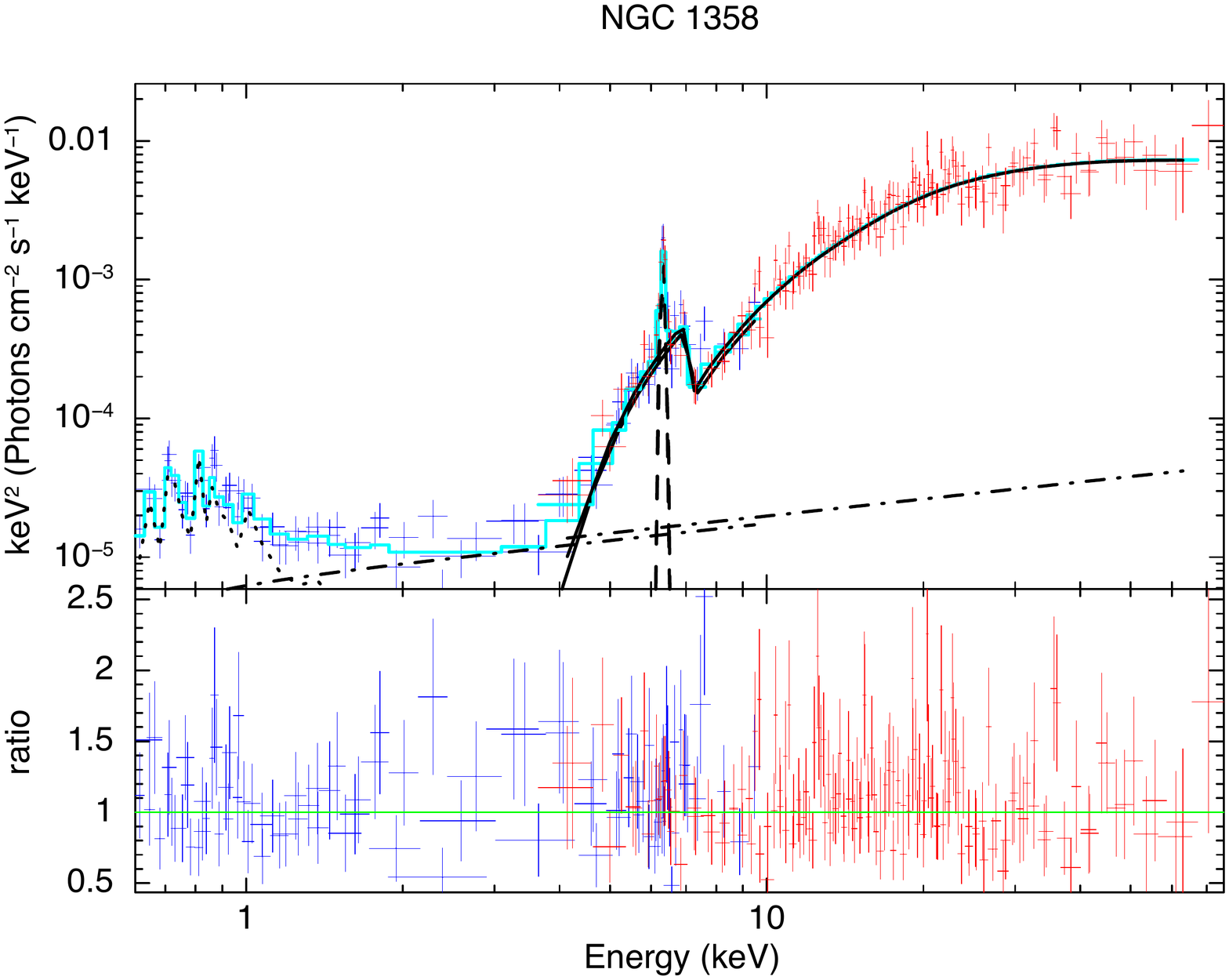} 
 \caption{Unfolded \XMM\ and \NuSTAR\ spectrum of NGC 1358 fitted with the \pexrav\ model (top) and ratio between data and model (bottom). The \XMM\ data is plotted in blue, while the \NuSTAR\ data is plotted in red. The best-fit models prediction is plotted as a cyan solid line. The single components of the model are plotted in black with different line styles, i.e., the absorbed intrinsic continuum as a solid line, the reflection component and Fe K$\alpha$ line as a dashed line, the scattered component as a dash-dotted line and the \textit{mekal} component as a dotted line.}
 \label{fig:pexrav}
 \end{figure}

We report in Table \ref{Table:best-fit} the best-fit parameters for the analysis of the joint \NuSTAR--\XMM\ spectra using model B. The photon index is $\Gamma$ = 1.59$_{-0.11}^{+0.11}$. The  best-fit column density is N$_{\rm H}$ = 0.76$_{-0.09}^{+0.09}$ $\times$ $10^{24}$\,cm$^{-2}$. In agreement with what we found using Model A, the source is near the threshold of CT-AGN. We present the unfolded \NuSTAR\ and \XMM\ spectrum of NGC 1358, fitted with the model B and ratio between data and model, in figure \ref{fig:pexrav}.

In summary, both phenomenological models suggest that obscuration is near the Compton-thick threshold, such that the source cannot be confirmed as CT-AGN at $>$3\,$\sigma$ confidence level. However, the photon indices obtained above are far from the typical value observed in AGN \citep[$\Gamma$ $\sim$1.8, see, e.g.,][]{marchesi2016}, showing that some components may not be well described by the above phenomenological models.
Therefore, more physically motivated models are needed to describe the spectra and extract the physical and geometrical properties of NGC 1358.

\begin{table*}
\renewcommand*{\arraystretch}{1.5}
\centering
\caption{Summary of Best-Fits of XMM-Newton and NuSTAR Data using Different Models}
\label{Table:best-fit}
\vspace{.1cm}
  \begin{tabular}{ccccccc}
       \hline
       \hline       
       Model&phenom&pexrav&MYTorus&MYTorus&MYTorus&borus02\\
       &&&(coupled)&(decoupled face on)&(decoupled edge on)&\\
       \hline
       $\chi^2$/dof&256/240&231/240&231/239&230/239&220/239&222/238\\
       $C_{Ins}$\footnote{$C_{Ins}$ = $C_{NuS/XMM}$ is the cross calibration between \NuSTAR\ and \XMM.}&1.06$_{-0.12}^{+0.14}$&1.13$_{-0.13}^{+0.14}$&1.12$_{-0.20}^{+0.15}$&1.13$_{-0.13}^{+0.15}$&1.17$_{-0.14}^{+0.16}$&1.16$_{-0.14}^{+0.12}$\\
       $\Gamma$&1.14$_{-0.12}^{+0.13}$&1.59$_{-0.11}^{+0.11}$&1.52$_{-*}^{+0.17}$&1.66$_{-0.26}^{+0.15}$&1.85$_{-0.23}^{+0.13}$&1.79$_{-0.04}^{+0.13}$\\
       N$\rm _H$\footnote{``line-of-sight'' column density in phenomenological models in $10^{24}$\,cm$^{-2}$}&0.95$_{-0.11}^{+0.11}$&0.76$_{-0.09}^{+0.09}$&...&...&...&...\\
       norm\footnote{normalization of components in different models at 1\,keV in photons\,keV$^{-1}$\,cm$^{-2}$\,s$^{-1}$.} 10$^{-2}$&0.03$_{-0.01}^{+0.02}$&0.04$_{-0.01}^{+0.02}$&0.13$_{-0.06}^{+0.16}$&0.20$_{-0.13}^{+0.24}$&1.61$_{-1.08}^{+1.26}$&1.26$_{-0.04}^{+0.79}$\\
        N$\rm _{H,eq}$&...&...&3.02$_{-1.12}^{+2.54}$&...&...&...\\
       $\theta\rm _{Tor}$\footnote{angle between the axis of the torus and the edge of torus in degree, where the covering factor $f_c$ = cos($\theta_{\rm Tor}$).}&...&...&...&...&...&84.0$_{-3.9}^{+*}$\\
       $\theta\rm _{Obs}$&...&..&62.53$_{-2.53}^{+4.33}$&...&...&87.1$_{-0.3}^{+*}$\\
       A$_S$&...&...&1.03$_{-0.51}^{+0.55}$&0.78$_{-0.26}^{+0.47}$&0.23$_{-0.06}^{+0.18}$&...\\
       N$\rm _{H,Z}$&...&...&...&1.19$_{-0.22}^{+0.27}$&2.40$_{-0.44}^{+0.40}$&2.40$_{-0.12}^{+0.39}$\\
       N$\rm _{H,S}$&...&...&...&5.25$_{-2.26}^{+*}$&0.50$_{-0.09}^{+0.12}$&0.65$_{-0.16}^{+0.05}$\\
       $f_s$ 10$^{-2}$&1.69$_{-0.51}^{+0.70}$&1.92$_{-0.57}^{+0.76}$&0.08$_{-0.08}^{+0.14}$&0.12$_{-0.12}^{+0.36}$&0.05$^{+0.02}_{-0.07}$&0.05$^{+0.01}_{-0.01}$\\
       kT\footnote{temperature in the thermal component \textit{mekal} in keV.}&0.49$_{-0.13}^{+0.09}$&0.49$_{-0.15}^{+0.10}$&0.58$_{-0.10}^{+0.07}$&0.57$_{-0.12}^{+0.07}$&0.58$_{-0.12}^{+0.06}$&0.52$_{-0.11}^{+0.06}$\\
       abund\footnote{abundance in the thermal component \textit{mekal}.}&0.05$_{-0.04}^{+0.08}$&0.07$_{-0.05}^{+0.19}$&0.03$_{-0.02}^{+0.04}$&0.04$_{-0.02}^{+0.05}$&0.11$_{-0.08}^{+18.29}$&0.05$_{-0.03}^{+0.08}$\\
       F$_{2-10}$\footnote{Flux between 2--10\,keV in $10^{-13}$ erg cm$^{-2}$ s$^{-1}$.}&4.18$_{-0.59}^{+0.23}$&4.09$_{-0.68}^{+0.26}$&4.03$_{-2.99}^{+100.}$&4.03$_{-3.21}^{+0.19}$&3.84$_{-1.75}^{+0.28}$&3.87$_{-3.86}^{+1.11}$\\
       F$_{10-40}$\footnote{Flux between 10--40\,keV in $10^{-12}$ erg cm$^{-2}$ s$^{-1}$.}&8.22$_{-1.56}^{+0.27}$&8.68$_{-0.94}^{+0.28}$&8.48$_{-8.48}^{+7.99}$&8.55$_{-3.77}^{+0.13}$&8.51$_{-2.48}^{+0.26}$&8.51$_{-8.51}^{+0.46}$\\
       L$_{2-10}$}\footnote{Intrinsic luminosity between 2--10\,keV in $10^{43}$ erg s$^{-1}$.&0.116$_{-0.004}^{+0.005}$&0.07$_{-0.01}^{+0.01}$&0.28$_{-0.02}^{+0.02}$&0.34$_{-0.02}^{+0.03}$&2.06$_{-0.11}^{+0.11}$&1.77$_{-0.11}^{+0.10}$\\
       L$_{10-40}$}\footnote{Intrinsic luminosity between 10--40\,keV in $10^{43}$ erg s$^{-1}$.&0.36$_{-0.02}^{+0.01}$&0.11$_{-0.11}^{+0.12}$&0.48$_{-0.03}^{+0.04}$&0.49$_{-0.03}^{+0.03}$&2.23$_{-0.12}^{+0.12}$&2.09$_{-0.12}^{+0.12}$\\
       \hline
	\hline
	\vspace{0.02cm}
\end{tabular}

\tablecomments{We summarize here the best-fits of joint \NuSTAR--\XMM\ spectra using different models referred in section \ref{sec:spectral}. We also report the statistics and degree of freedom for each fit.}
\end{table*}

\subsection{Physical Models}\label{sec:physical}
\subsubsection{\MYTorus}\label{section:MYTorus}
The first physically motivated model applied in our analysis is \MYTorus\ \citep{MYTorus2009,MYTorus2012,MYTorus2015}. The basic geometry of \MYTorus\ model consists of a torus that has a fixed half-opening angle, $\theta$ = 60$^{\circ}$, with a circular cross section. 

An advantage of the physically motivated \MYTorus\ model is that the main components observed in the spectrum of an obscured AGN can be treated self-consistently. More in detail, \MYTorus\ model is composed of three parts: the direct continuum, the Compton-scattered component and fluorescent lines. 
The direct continuum, which is also called zeroth-order continuum, is the ``line-of-sight'' observed continuum, i.e., the intrinsic X-ray continuum as observed after the absorption caused by the torus. In \MYTorus, this first component is a multiplicative factor (a multiplicative table in \XSPEC), which is applied to the intrinsic continuum. 
The second component is the scattered continuum and it models those photons that are Compton-scattered into the observer ``line-of-sight'' by the gas in the environment of the SMBH. If the covering factor of the torus differs significantly from the fixed \MYTorus\ value, $f_c$ = cos($\theta$) = 0.5, or if there is a non-negligible time delay between the intrinsic continuum emission and the Compton-scattered continuum one, i.e., the center region is not compact and the intrinsic emission varies rapidly, the two components could have different normalizations. To take these effects into account, the scattered continuum is multiplied by a relative normalization, which is noted as $A_S$ \citep{MYTorus2012}.
Finally, the third component models the most prominent fluorescent lines, i.e., the Fe K$\alpha$ and Fe K$\beta$ lines, at 6.4\,keV and 7.06\,keV, respectively. Analogously to $A_S$, the relative normalization between the fluorescent lines and direct continuum is noted as $A_L$. 
In \XSPEC, $A_S$ and $A_L$ are implemented as two \textit{constant} components before the additive tables, while the normalization of the three components are set to be the same. Following previous works, the two relative normalizations are set to be equal, i.e., $A_S$=$A_L$. 

In \XSPEC\ our model is described as follows:
\begin{equation}
\label{eq:coupled}
\begin{aligned}
ModelC=&constant_1*phabs*\\
&(mytorus\_ Ezero\_v00.fits*zpowerlw\\
&+A_S*mytorus\_scatteredH500\_v00.fits\\
&+A_L*mytl\_V000010nEp000H500\_v00.fits\\
&+constant_2*zpowerlw+mekal)\\
\end{aligned}
\end{equation}
where table \textit{mytorus\_Ezero\_v00.fits} is the zeroth-order continuum component, \textit{mytorus\_scatteredH500\_v00.fits} accounts for the scattered continuum, and \textit{mytl\_V000010nEp000H500\_v00.fits} models the fluorescent lines.

The \MYTorus\ model can be used in two different configurations, named `coupled' and `decoupled' \citep{MYTorus2012}. We test both of them on our data, and we report the results in the following sections.

\subsubsection{\MYTorus\ in `coupled' configuration}
In \MYTorus, the angle between the axis of the torus and the ``line-of-sight'', the so-called ``torus inclination angle'', is a free parameter, that we hereafter define as $\theta_{\rm obs}$. The inclination angle varies in the range $\theta_{\rm obs}$=[0--90]$^\circ$, where $\theta_{\rm obs}$=0$^\circ$ models a torus observed ``face-on'', and  and $\theta_{\rm obs}$=90$^\circ$ is observed ``edge-on''. In the `coupled' configuration, $\theta_{\rm obs}$ is set to be the same for all three \MYTorus\ components.

We report in Table \ref{Table:best-fit} the best-fit parameters obtained using the \MYTorus\ `coupled' model. We fit the \XMM\ data in the 0.6--10\,keV energy range to avoid a known \MYTorus\ fit issue below 0.6\,keV, that may cause large statistical errors (more details are available in the \MYTorus\ manual\footnote{http://mytorus.com/mytorus-instructions.html}). The best-fit photon index is $\Gamma$ =  1.52$_{-*}^{+0.17}$ (the lower limit of the photon index cannot be constrained in our modeling since it falls below 1.4, the value that is the smallest value that can be tested with \MYTorus). The photon index of the Compton-scattered continuum and of the iron emission feature component set to be the same with that of direct continuum. The equatorial column density is N$_{\rm H,eq}$ = 3.02$_{-1.12}^{+2.54}$$~\times~10^{24}$\,cm$^{-2}$. The inclination angle is $\theta_{\rm obs}$ = 62.53$_{-2.53}^{+4.33}$$^\circ$, suggesting that we are observing through the `brink' of the torus. The ``line-of-sight'' column density is defined as N$_{\rm H,l.o.s.}$ = N$_{\rm H,eq}$ [1 - 4 cos$^2\theta_{\rm obs}$]$^{1/2}$ = 1.17$_{-1.17}^{+3.94}$$~\times~10^{24}$\,cm$^{-2}$. The reduced $\chi^2$ is $\chi^2_\nu$ = 231/239 = 0.97.

\subsubsection{\MYTorus\ in `decoupled' configuration}\label{sec:myt_dec}
The `decoupled' \MYTorus\ model, which is first introduced in \citet{MYTorus2012}, adds flexibility to the \MYTorus\ model as it allows the users to model the absorber's structure with a more general geometry and even simulate a clumpy distribution of the obscuring material. In this configuration,
the direct continuum is a pure ``line-of-sight" quantity, and the inclination angle of the direct continuum is fixed to $\theta_{\rm obs,Z}$ = 90$^\circ$, such that the column density of the direct continuum models the ``line-of-sight" column density, N$_{\rm H,Z}$.
The inclination angle of the Compton-scattered continuum and fluorescent lines is instead set to be either observed ``face-on'', such that $\theta_{\rm obs,S,L}$ = 0$^\circ$, or observed ``edge-on'', i.e., $\theta_{\rm obs,S,L}$ = 90$^\circ$. The ``face-on'' configuration mimics the reprocessed emission coming from the ``back-side'' of the torus, which is expected to be more prominent in a patchy, less uniform torus, where the photons emitted by the ``back-side'' of the torus have a smaller chance of being absorbed before reaching the observer. In the ``edge-on'' scenario, instead, the photons are reprocessed by the obscuring material lying between the AGN and the observer, and an ``edge-on--dominated'' reprocessed emission therefore favors a more uniform distribution of the obscuring material.
In the `decoupled' \MYTorus\ model, the column density of the scattered continuum and of the fluorescent lines, N$_{\rm H,S}$, describes the ``global average'' column density of the torus, i.e., the average column density of the obscuring material, which can significantly differ from the ``line-of-sight'' value in an inhomogeneous, patchy torus. 

In \citet{MYTorus2012}, the `decoupled' \MYTorus\ model is used adding to the model both the $\theta_{\rm obs,S,L}$ = 90$^\circ$ and the $\theta_{\rm obs,S,L}$ = 0$^\circ$ reprocessed components, thus we first test this model, where both components contribute to the total reprocessed emission. In such a scenario, the best-fit is $\chi^2$/d.o.f = 222/238 = 0.93, while the intensity of the ``face-on'' component is 15 times smaller than the intensity of the ``edge-on'' component, suggesting that the reprocessed emission in NGC 1358 comes mostly from material located between the AGN and the observer. For this reason, following the approach described in \citet{MYTorus2015}, we re-fit our data twice, each time using only one of the two reprocessed component configurations. The best-fit for the pure ``back-side'' reflection model, i.e., $\theta_{\rm obs,S,L}$ = 0$^\circ$ is presented in Table \ref{Table:best-fit}. The photon index is $\Gamma_{\rm \theta,S=0}$ = 1.66$_{-0.26}^{+0.15}$, the ``line-of-sight'' column densities is N$_{\rm H,Z,\theta,S=0}$ = 1.19$_{-0.22}^{+0.27}$ $\times$ $10^{24}$\,cm$^{-2}$ and the ``global average'' column density is N$_{\rm H,S,\theta,S=0}$ = 5.25$_{-2.26}^{+*}$ $\times$ 10$^{24}$\,cm$^{-2}$. The best-fit by using only the $\theta_{\rm obs,S,L}$=90$^\circ$ is also presented in Table \ref{Table:best-fit}. The photon index is $\Gamma_{\rm \theta,S=90}$ = 1.85$_{-0.23}^{+0.13}$. The ``line-of-sight" column densities is N$_{\rm H,Z,\theta,S=90}$ = 2.40$_{-0.44}^{+0.40}$ $\times$ $10^{24}$\,cm$^{-2}$. The ``global average" column density is N$_{\rm H,S,\theta,S=90}$ = 0.50$_{-0.09}^{+0.12}$ $\times$ $10^{24}$\,cm$^{-2}$. The ``global average" column density is a few times smaller than the ``line-of-sight" column density, suggesting a patchy torus scenario, where the AGN is observed through an over-dense cloud. 

\begin{figure}[htpb]
\centering
\includegraphics[width=.5\textwidth]{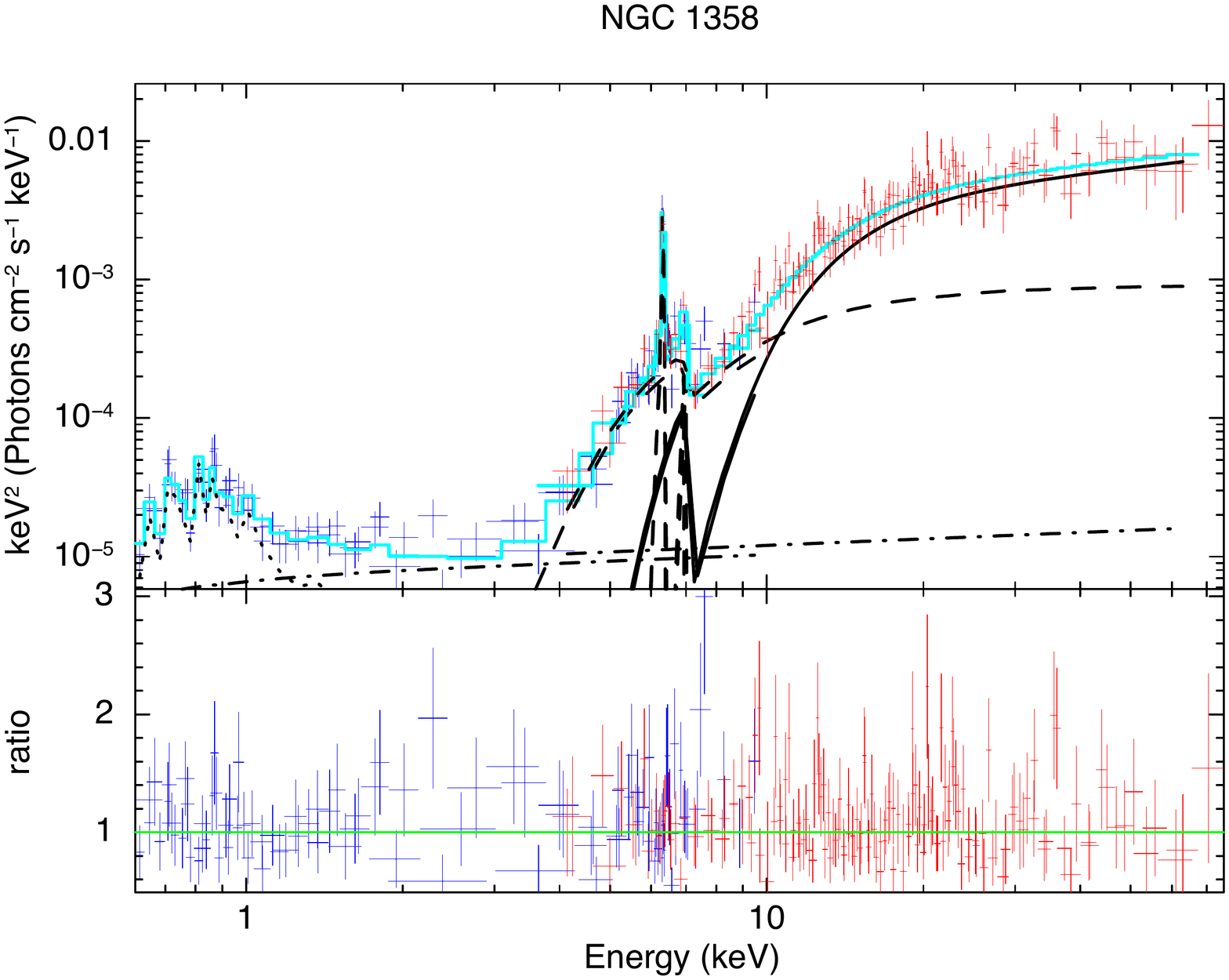} 
 \caption{Unfolded \XMM\ and \NuSTAR\ spectrum of NGC 1358, fitted with the `decoupled' \MYTorus\ model in ``edge on'' configuration (top) and ratio between data and model (bottom). The \XMM\ data is plotted in blue, while the \NuSTAR\ data is plotted in red. The best-fit model prediction is plotted as a cyan solid line. The single components of the model are plotted in black with different line styles, i.e., the absorbed intrinsic continuum as a solid line, the Compton-scattered continuum and the fluorescent lines as a dashed line, the scattered component as a dash-dotted line and the \textit{mekal} component as a dotted line.}
 \label{fig:decoupled90}
 \end{figure}

In conclusion, the best-fit of `decoupled' \MYTorus\ model in both ``face-on'' and ``edge-on'' configurations confirm the Compton-thick origin of NGC 1358 at 3\,$\sigma$ confidence level. The \MYTorus\ `decoupled' model in ``edge on'' configuration produced the best-fit, $\chi^2_\nu$ = $\chi^2$/d.o.f = 220/239 = 0.92 and most reasonable photon index \citep[$\Gamma$ $\sim$1.8, see, e.g.,][]{marchesi2016}, among all the models and is thus our favorite model. Figure \ref{fig:decoupled90} shows the unfolded \NuSTAR\ and \XMM\ spectrum of NGC 1358, fitted with the `decoupled' \MYTorus\ model in ``edge on'' configuration.

\subsubsection{BORUS02}
While \MYTorus\ is known to be effective in modeling the X-ray spectra of heavily obscured AGN, it assumes a fixed torus opening angle ($\theta_{\rm Tor}$ = 60$^\circ$, i.e., a covering factor $f_c$ = cos $\theta_{\rm Tor}$ = 0.5), limiting the model to a single torus geometry in `coupled' mode and do not allowing one to directly measure the covering factor even in the `decoupled' mode, although the latter can be in principle used to mimic different geometries of the obscuring material. 
To complement our analysis, we therefore fit the NGC 1358 spectrum using the recently published \borus\ model \citep{Borus}, an updated version of the so-called \bntorus\ model \citep{BNtorus}. In \borus\ the torus covering factor is a free parameter varying in range of $f_c$ = [0.1--1], corresponding to a torus opening angle $\theta_{\rm Tor}$ = [0--84]$^\circ$. 

\borus\ is used in the following \XSPEC\ configuration:
\begin{equation}
\label{eq:Borus}
\begin{aligned}
ModelD=&constant_1*phabs*(borus02\_v170323a.fits\\
&+zphabs*cabs*zpowerlw+constant_2*\\
&zpowerlw+mekal) 
\end{aligned}
\end{equation}
where $borus02\_v170323a.fits$ is an additive table that models the reprocessed components, including the fluorescent line emission and the reprocessed continuum. The ``line-of-sight'' absorption is modeled by the $zphabs\times cabs$ including Compton scattering lost out of the ``line-of-sight''. that includes the effect of Compton scattering. The other components are similar to \MYTorus.

\begin{figure}[htpb]
\centering
\includegraphics[width=.5\textwidth]{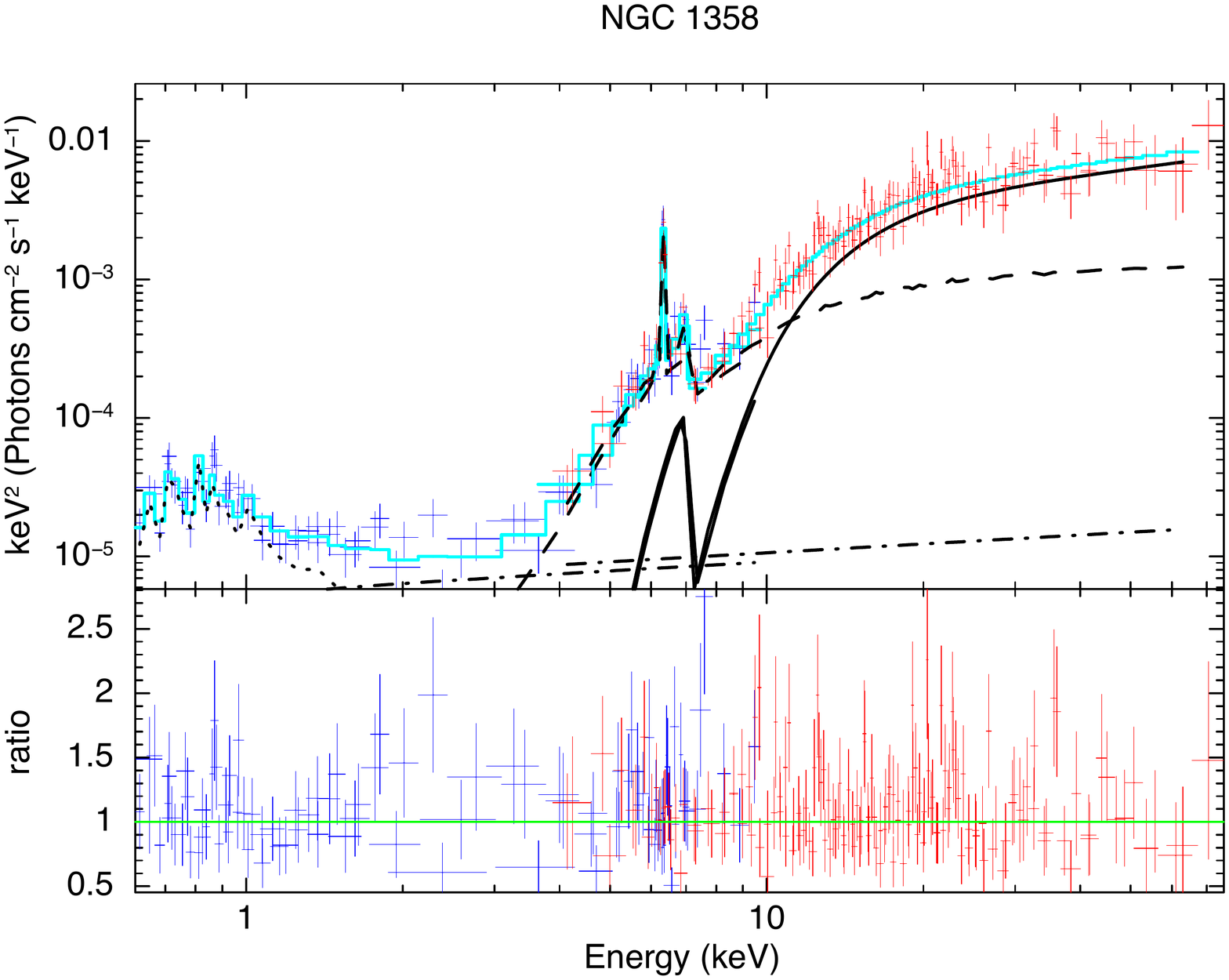} 
 \caption{Unfolded \XMM\ and \NuSTAR\ spectrum of NGC 1358, fitted with the \borus\ model (top) and ratio between data and model (bottom). The \XMM\ data is plotted in blue, while the \NuSTAR\ data is plotted in red. The best-fit model prediction is plotted as a cyan solid line. The single components of the model are plotted in black with different line styles, i.e., the absorbed intrinsic continuum as a solid line, the reflection component and fluorescent lines as a dashed line, the scattered component as a dash-dotted line and the \textit{mekal} component as a dotted line.}
 \label{fig:borus}
 \end{figure}

The best-fit results are presented in Table \ref{Table:best-fit}. The photon index is $\Gamma$ = 1.79$_{-0.04}^{+0.13}$. The ``line-of-sight" column density is N$\rm _{H,Z}$ = 2.40$_{-0.12}^{+0.39}$ $\times~10^{24}$\,cm$^{-2}$, while the column density of the torus is N$\rm _{H,S}$ = 0.65$_{-0.16}^{+0.05}$ $\times~10^{24}$\,cm$^{-2}$, which is in good agreement with the results of the \MYTorus\ `decoupled' model in ``edge on'' configuration. The covering factor of the torus is $f_{c}$ $<$0.17, i.e., a disk-like torus, which is strongly different from the set-up of \MYTorus\ model, which will be further discussed in section \ref{discussion}. Finally, the angle between the torus axis and the observer ``line-of-sight'' is $\theta_{\rm obs}$ $>$86.8$^\circ$, suggesting that the `edge on' scenario is favored. Figure \ref{fig:borus} shows the unfolded \XMM\ and \NuSTAR\ spectrum of NGC 1358, fitted by the \borus\ model.

\subsection{Summary of the spectral fit results}
Based on the fit statistic and on the reliability of the best-fit parameters, we believe that the `decoupled' \MYTorus\ in ``edge-on'' configuration is the best-fit model for NGC 1358. In our fits, all models have good statistic, with $\chi_\nu$ $\sim$0.9--1.1, nonetheless their physical interpretation varies significantly. For example, model B suggests a reflection dominated scenario; in both the `coupled' \MYTorus\ and `decoupled' \MYTorus\ in ``face-on'' configuration, the Compton-scattered component contribution to the total observed emission is as significant as the direct continuum one; finally, the `decoupled' \MYTorus\ in ``edge-on'' configuration, which gives the best statistic, suggests that the direct continuum dominates at E$>$10\,keV and the contribution of the reprocessed component is relatively smaller, and a similar result is also obtained using \borus.

 We are going to use `decoupled' \MYTorus\ in ``edge-on'' configuration as a reference in the rest of the paper because it allows one to compute the iron K$\alpha$ line equivalent width, which cannot be done as straightforwardly using \borus, whose iron K$\alpha$ line and the reprocessed component are coupled together. The only exception will be in Section \ref{sec:cf}, where the results of the \borus\ model, in which $f_c$ is a free parameter, will be used in discussing the torus covering factor; however, as mentioned above, the best-fit results of \borus\ are in full agreement with those of the `decoupled' \MYTorus\ in ``edge-on'' configuration.
\subsection{Flux and Column Density Variability}\label{sec:variability}
While the majority of tori models, such as \MYTorus\ and \borus, assume a uniform distribution of the obscuring material, several works in the last 25 years have shown that a clumpy distribution of optically thick clouds, the so-called ``patchy torus'', is in fact a scenario favored by the observations \citep[see, e.g.,][]{Krolik1988,Antonucci1993,Jaffe2004,Elitzur06,Nenkova08}. 

\citet{Mendoza15} study the silicates spectral features at 10 and 18\,$\mu$m in MIR band, showing that the distribution of the obscuring material around the SMBH has a clumpy structure. Following this scenario, it may be possible to observe changes of the ``line-of-sight" column density, N$\rm _{H,Z}$, with time. For this purpose, we divide our \NuSTAR\ and \XMM\ observations into several shorter observations and extract a spectrum for each of them having at least $\sim$150 net counts (except for the first pn observation, which has 75 net counts, due to a background flare at the beginning of the \XMM\ observation).

\begin{figure}[htpb]
\centering
\includegraphics[width=.45\textwidth]{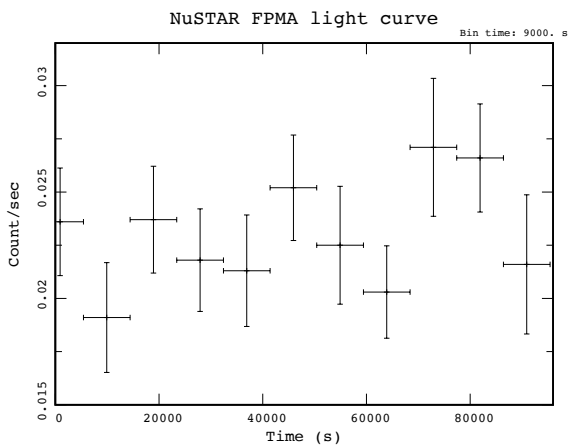} 
 \caption{Background subtracted light curve of \NuSTAR\ module FPMA. The bin time equals to 9 ks. The average count rate is r = 2.3$\pm$0.3$\times$10$^{-2}$\,cts s$^{-1}$.}
 \label{fig:lightcurve}
 \end{figure}

\begin{figure}[htpb]
\centering
\includegraphics[width=.45\textwidth]{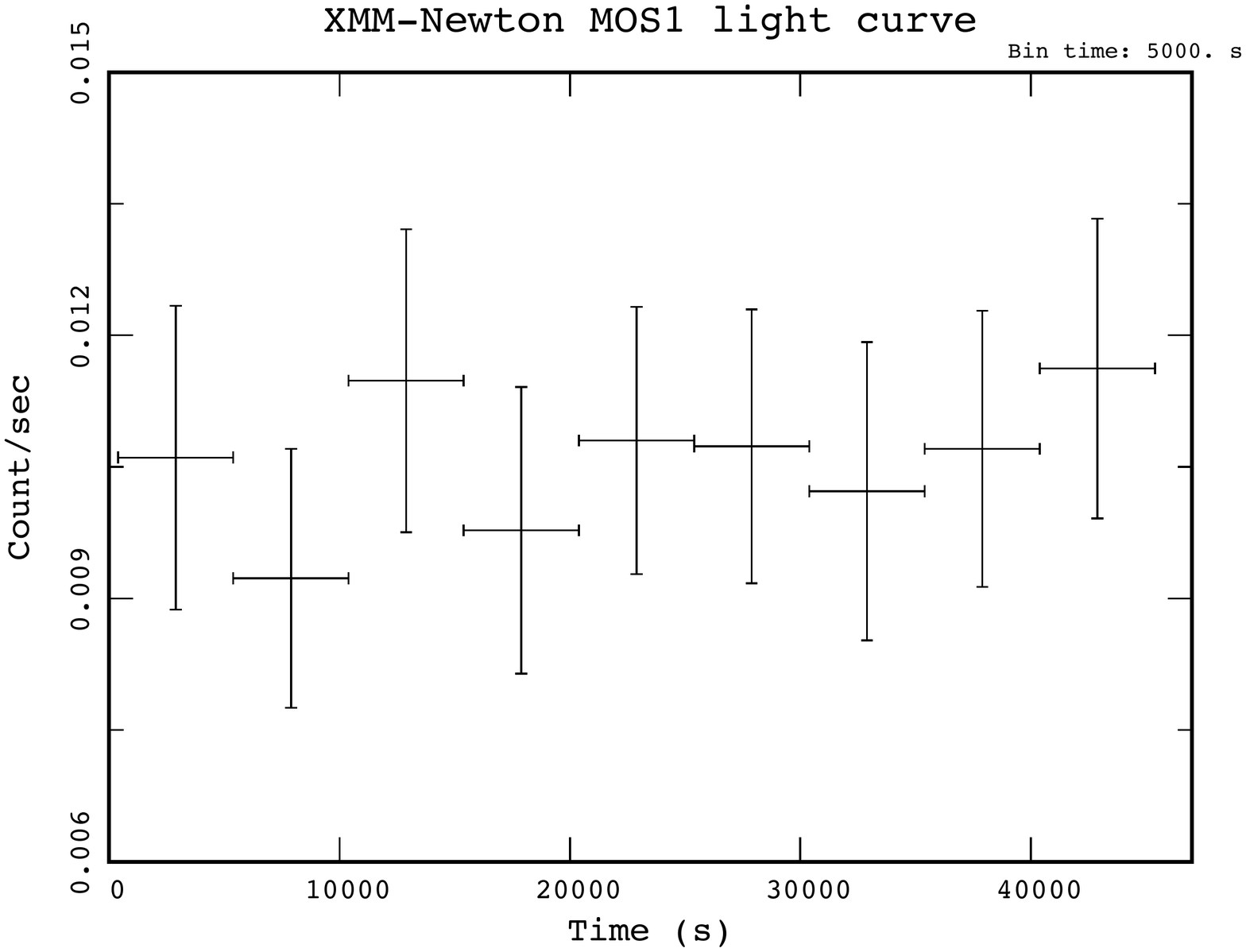} 
 \caption{Background subtracted light curve of \XMM\ MOS1. The bin time equals to 5 ks. The average count rate is r = 9.8$\pm$1.6$\times$10$^{-3}$\,cts s$^{-1}$.}
 \label{fig:m1_lc}
 \end{figure}

Since the \XMM\ observation started $\sim$13 hours after the \NuSTAR\ one, we divide our \NuSTAR\ and \XMM\ observations into three blocks: $i$) The first one contains the 26\,ks of \NuSTAR\ observation taken before the beginning of the \XMM\ observation. $ii$) The second and third ones are obtained dividing the remaining 24\,ks of \NuSTAR\ data and the 48\,ks \XMM\ data in two even pieces, each one includes 12\,ks \NuSTAR\ observation and 24\,ks of \XMM\ observation. We remind that the \NuSTAR\ observation was taken in blocks of $\sim$3\,ks each, therefore the \NuSTAR\ and \XMM\ observations start at different time but finish at the same time. The background subtracted \NuSTAR\ FPMA and \XMM\ MOS1 light curves of NGC 1358 are shown in Figure \ref{fig:lightcurve} and Figure \ref{fig:m1_lc}.

\begin{table*}

\renewcommand*{\arraystretch}{1.5}
\centering
\caption{Physical properties in different time ranges. Fit performed with phenomenological model}
\label{Table:variability}
\vspace{.1cm}
  \begin{tabular}{ccccc}
       \hline
       \hline     
       Observation&Overall observations&0-26\,ks NuSTAR&26-38\,ks NuSTAR \& 0-24\,ks XMM&38-50\,ks NuSTAR \& 24-48\,ks XMM\\
       \hline
       $\chi^2$/dof&256/240&84/94&92/87&66/85\\
       $\Gamma$&1.14$^{+0.12}_{-0.13}$&1.48$^{+0.24}_{-0.25}$&1.30$_{-0.21}^{+0.23}$&1.27$_{-0.25}^{+0.26}$\\
       N$_H$\footnote{units of all column density are $10^{24}$\,cm$^{-2}$}&0.95$_{-0.11}^{+0.11}$&1.41$_{-0.29}^{+0.30}$&1.15$_{-0.18}^{+0.20}$&1.06$_{-0.18}^{+0.19}$\\
	F$_{-10~\rm keV}$\footnote{2-10 keV observe flux 10$^{-13}$ erg cm$^2$ s$^{-1}$}&4.18$_{-0.59}^{+0.23}$&5.15$_{-2.36}^{+0.46}$&4.14$_{-2.17}^{+0.38}$&4.39$_{-1.71}^{+0.48}$\\
       \hline
       \hline
	\vspace{0.06cm}
\end{tabular}
\end{table*}

The three spectra are fitted with the phenomenological model, since we are mostly interested in measuring flux and/or column density variation. We report in Table \ref{Table:variability} the best-fit $\Gamma$, N$\rm _H$ and flux of 2--10\,keV for each of the three sub-observations. The flux and column density measured in the different blocks are consistent with each other, although those of block 1 are marginally offset with respect to the other two, a result that is mostly due to a calibration offset between \XMM\ and \NuSTAR. In fact, when fitting the block 2 and 3 \NuSTAR\ data alone we find a smaller discrepancy and the results are in agreement at 90\% confidence level.

\begin{figure}[htpb]
\centering
\includegraphics[width=.5\textwidth]{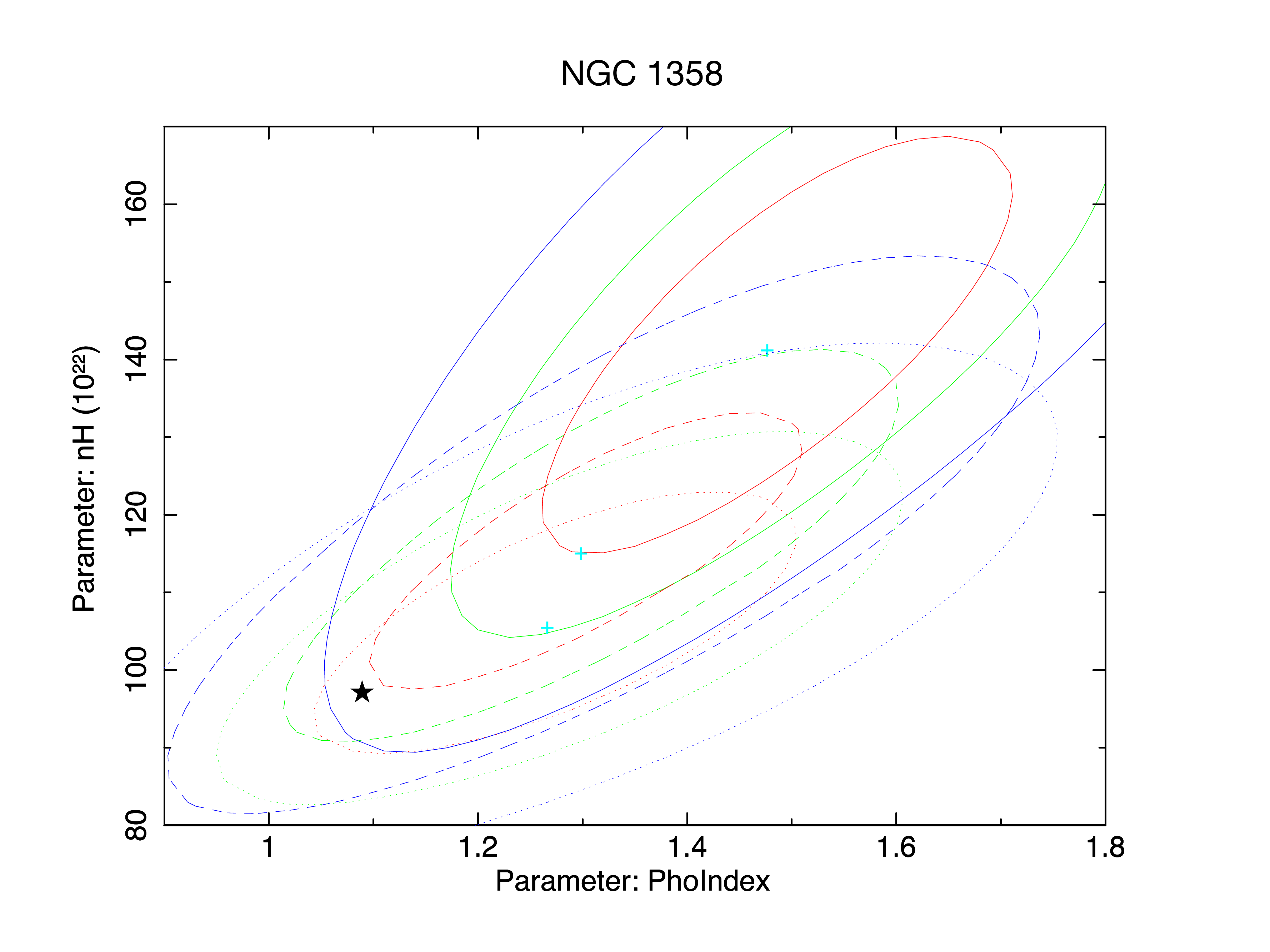} 
 \caption{Contour of photon index and column density of three time blocks fitted with phenomenological model. The red, green and blue lines are at 68\%, 90\% and 99\% confidence level. Contours of time blocks 1 to 3 are solid, dash and dot lines respectively. The photon index and column density of the overall observations is marked as a black star.}
 \label{fig:contour}
 \end{figure}

We also plot the contour of photon index and column density in Figure \ref{fig:contour}, where the red, green and blue lines are at 68\%, 90\% and 99\% confidence level. Time blocks 1 to 3 are plotted in solid, dash and dot lines respectively. As can be seen, both quantities are consistent among the three blocks, at a 90\% confidence level.

Finally, we fit the \NuSTAR\ 3--79\,keV light curve (Figure \ref{fig:lightcurve}) with a constant corresponding to the average count rate of our source, which is r = 2.3$\pm{0.3}$$\times$10$^{-2}$\,cts\,s$^{-1}$. The best-fit statistic is $\chi^2$ = 9.3, and the fit has 10 degrees of freedom. At 99\% confidence level, the light curve is strongly different from a constant if $\chi^2$ $>$$23.2$. We repeat the above process for the 0.6--10\,keV background subtracted light curve of \XMM\ MOS1 module (Figure \ref{fig:m1_lc}). The average count rate is r = 9.8$\pm{1.6}$$\times$10$^{-3}$\,cts\,s$^{-1}$ and the best-fit statistic is $\chi^2$ = 19.0 with 9 degrees of freedom. At 99\% confidence level, the light curve is strongly different from a constant if $\chi^2$ $>$$21.7$.  Therefore, we find no obvious evidence of variability in either flux of absorbing column density.

\subsection{Equivalent width of the iron K$\alpha$ line}
Thanks to the excellent count statistics provided by \NuSTAR\ and \XMM\ in the 5--8\,keV band, we were also able to place strong constraints on the Fe K$\alpha$ line equivalent width (EW), a significant improvement with respect to \citet{marchesi2017APJ}, where only an upper limit on EW could be derived. We measure equivalent width EW$_{\rm pow}$ = 0.72$_{-0.16}^{+0.16}$\,keV and EW$_{\rm pex}$ = 0.63$_{-0.14}^{+0.15}$\,keV using the model A and model B respectively. 

To measure the Fe K$\alpha$ line EW with \MYTorus\ we use the approach described in \citet{MYTorus2015}. We therefore first measure the continuum flux, without including the emission line, at E$_{\rm K\alpha}$ = 6.4 keV. We then compute the flux of the fluorescent lines component in the energy range E = [0.95\,E$_{\rm K\alpha}$--1.05\,E$_{\rm K\alpha}$], i.e., between 6.08 and 6.72\,keV, rest-frame. EW is then computed by multiplying by (1 + \textit{z}) the ratio between the fluorescent line flux and the monochromatic continuum flux.
We obtain EW$_{\rm coupl}$ = 0.69$_{-0.12}^{+0.13}$\,keV, EW$_{\rm decoupl,\theta=90}$ = 0.70$_{-0.11}^{+0.14}$\,keV and EW$_{\rm decoupl,\theta=0}$ = 0.65$_{-0.13}^{+0.12}$\,keV, such that all the \MYTorus\ EW values are in good agreement with those obtained in the phenomenological model.

%
%

\section{Discussion and Conclusions}
\label{discussion}

\subsection{Compared with previous results}
In this work, we report the results of the spectral analysis of quasi-simultaneous \NuSTAR\ (50\,ks) and \XMM\ (48\,ks) observations of NGC 1358, a nearby Seyfert 2 galaxy, which was recently found to be a CT-AGN candidate \citep{marchesi2017APJ} based on its combined \textit{Swift}-BAT and \cha\ spectrum. The limited quality of the \cha\ and \textit{Swift}-BAT spectra was reflected in the rather large ($\sim$30--40\%) 90\% confidence parameter uncertainties, and NGC 1358 Compton thickness could not be validated beyond the 1\,$\sigma$ confidence level. More in detail, \citet{marchesi2017APJ} used \MYTorus\ `coupled' configuration to model the combined \cha--\textit{Swift}-BAT spectrum of NGC 1358. Their best-fit ($\chi^2$/d.o.f. = 14.8/13) gave a column density of N$_{\rm H,Z}$ = 1.05$_{-0.36}^{+0.42}$ $\times$ 10$^{24}$\,cm$^{-2}$ assuming the inclination angle to be $\theta_{\rm obs}$ = 90$^\circ$. 

NGC 1358 was first reported as a candidate CT-AGN by \citet{Marinucci12}, which measured the X-ray spectral properties of NGC 1358 by fitting a 2-10 keV \XMM\ spectrum observed in 2005 (exposure time 12.7\,ks)  with a Compton-reflection model (\pexrav). The column density they obtained is N$_{\rm H,Z}$ = 1.30$_{-0.60}^{+8.50}$ $\times$ 10$^{24}$\,cm$^{-2}$, in good agreement with the findings of \citet{marchesi2017APJ}. Once again, the CT- nature of NGC 1358 was not confirmed at a $>$1\,$\sigma$ significance level.

The first main result of our analysis is therefore that NGC 1358 is a confirmed, \textit{bona-fide} Compton-Thick AGN, based on the two models which provide the most reliable fit, i.e., the `decoupled' \MYTorus\ model in ``edge on" configuration and \borus\ model. More in detail, our best-fit ``line-of-sight" column density obtained by the \MYTorus\ `decoupled' model in ``edge on" configuration is N$_{\rm H,Z}$ = 2.40$_{-0.44}^{+0.40}$ $\times~10^{24}$\,cm$^{-2}$ and for most of the parameters, the uncertainties are $<$20\% at 90\% confidence. 

\subsection{Intrinsic X-ray luminosity}
In table \ref{Table:best-fit}, we report the observed flux and intrinsic luminosity of NGC 1358 in the 2--10\,keV and 10--40\,keV energy ranges, for all the different models discussed in Section \ref{sec:spectral}. The observed flux values are consistent among both the phenomenological and the physical models, in both energy ranges. We instead observe significant differences between the best-fit luminosity values, since the `decoupled' \MYTorus\ model in ``edge-on'' configuration and the \borus\ model give a 2--10\,keV intrinsic luminosity L$\rm _{2-10\,keV}$ $\sim$2.06 $\times$ 10$^{43}$ erg s$^{-1}$, while the intrinsic luminosity in 2--10\,keV for the other models is L$\rm _{2-10\,keV}$ $<$0.4 $\times$ 10$^{43}$ erg s$^{-1}$, which is at least 6 times smaller than the best-fit models. This large difference is a direct consequence of the fits results: in the phenomenological models, the ``coupled'' \MYTorus\ model and the ``decoupled'' \MYTorus\ `face-on' model, NGC 1358 is intrinsically less luminous, has a harder (somehow unphysical, particularly in the phenomenological model) photon index, is less obscured and has a strong reprocessed component. In the ``decoupled'' \MYTorus\ `edge-on' model and in \borus, instead, NGC 1358 is more luminous, more obscured, has a softer, more typical photon index, and the reprocessed component is relative small.

Moreover, the intrinsic X-ray luminosity can also be inferred from luminosities derived at different wavelengths, e.g.,  the middle infrared (MIR) luminosity \citep[see, e.g.,][]{Elvis1978} and the [OIII] luminosity \citep[see, e.g.,][]{Heckman05}. The MIR luminosity of NGC 1358 is obtained by using the flux at 12\,$\mu$m, $F_{\rm 12\mu m}$ = 1.43$\pm$0.03 $\times$ 10$^{-2}$\,Jy \citep{wright10}: the corresponding luminosity is $L_{\rm 12\mu m}$ = 1.23 $\pm$ 0.03 $\times$ 10$^{42}$\,erg s$^{-1}$. Using the L$\rm _{MIR}$-L$\rm_{2-10\,keV}$ relation in \citet{Asmus15}, we then obtain the MIR-inferred X-ray intrinsic luminosity L$\rm_{2-10\,keV,MIR}$ $\sim$5.2$\times$ 10$^{41}$ erg s$^{-1}$. The [OIII] luminosity of NGC 1358 is reported in \citet{Whittle1992} and is L$\rm_{[OIII]}$ = 9.8 $\times$ 10$^{40}$ erg s$^{-1}$. Applying the L$\rm_{[OIII]}$-L$\rm_{2-10\,keV}$ relation from \citet{Georgantopoulos10}, we obtain the [OIII]--inferred X-ray intrinsic luminosity L$\rm_{2-10\,keV,[OIII]}$ $\sim$5.31$\times$ 10$^{42}$ erg s$^{-1}$. Notably, these luminosities are in better agreement with those computed using the phenomenological models, the `coupled' \MYTorus\ model and the `decoupled' \MYTorus\ model in ``face-on'' configuration, rather than with that derived using the `decoupled', ``edge-on'' \MYTorus\ model that we selected as our best-fit model. However, we point out that both L$\rm _{MIR}$-L$\rm_{2-10\,keV}$ and the L$\rm_{[OIII]}$-L$\rm_{2-10\,keV}$ relations are derived using phenomenological models, which are known to be less reliable than the physical models when fitting the heavily obscured AGN, such as NGC 1358. The low MIR luminosity reported above may be due to the low covering factor of the torus in NGC 1358 (see next section). However, despite being disfavored by the data, the possibility that the source is less obscured and intrinsically fainter than suggested by the best-fit model cannot be ruled out.

\subsection{Covering factor}\label{sec:cf}
One of the advantages in fitting the X-ray spectra of heavily obscured AGNs using a physical model is the possibility to measure specific tori parameters, such as the torus covering factor, $f_c$. In \borus\ the covering factor is a free parameter: we find that the best-fit \borus\ solution supports a low--$f_c$ scenario for NGC 1358, with  $f_c$ $<$0.17, thus suggesting a ``disk-like" torus. Such an evidence is supported also by the \MYTorus\ best-fit model: in fact, while in \MYTorus\ $f_c$ is not a free parameter, the ratio between the normalization of the intrinsic continuum and the one of the Compton-scattered continuum, $A_S$, can be used as a proxy of the torus covering factor. More in detail, from our best-fit \MYTorus\ model, i.e., the `decoupled edge on' one, we obtain $A_S$ = 0.23$_{-0.06}^{+0.18}$. The estimated covering factor is then $f_c$ = 0.5 $\times$ $A_S$ = 0.12$_{-0.03}^{+0.09}$, again supporting a ``disk-like " torus scenario. However, it is worth mentioning that \citet{MYTorus2012} points out that the existence of a non-negligible time delay between the two components can also result in the relative normalization being away from unity.

Finally, we checked for further evidence of a ``disk-like'' torus scenario using the ratio of the torus luminosity to the AGN luminosity as a proxy of the torus covering factor \citep{Stalevski16}. We indirectly infer the torus luminosity from the 12\,$\mu$m luminosity, which is dominated by the emission reprocessed by the torus. We can then derive a first order estimate of the dust torus covering factor, $f_c$ = $L_{\rm tor}$/$L_{\rm bol}$ $\sim$0.005, where $L_{\rm tor}$ is the torus luminosity and L$\rm _{bol}$ = 2.34 $\times$ 10$^{44}$\,erg\,s$^{-1}$ \citep{Woo02} is the bolometric luminosity of the AGN. While this result needs to be validated by a more accurate modeling and fit of the torus SED, it still points to a ``disk-like" torus scenario.

\subsection{Clumpiness of the torus}
In section \ref{sec:physical}, we computed both the ``line-of-sight" column density, N$\rm _{H,Z}$, and the torus ``global average" column density, N$\rm _{H,S}$, using both \MYTorus\ in its `decoupled' configuration, and \borus. In both cases, we find a mild ($\sim$4 times) difference between the ``line-of-sight" column density and the torus ``global average" column density, an evidence which supports an inhomogeneous distribution of the obscuring material surrounding the accreting SMBH in NGC 1358. We point out that this statement may seem in contradiction with what we stated in Section \ref{sec:myt_dec} when presenting the \MYTorus\ decoupled ``face-on'' model, i.e., the fact that the ``face-on'' model, rather than the ``edge-on'' one, should be more effective in characterizing a patchy torus scenario, where there is a higher chance to observe the reprocessed emission coming from the ``back-side'' of the torus.  However, this potential discrepancy can be explained assuming that in NGC 1358 the torus covering factor is indeed small, and the reprocessed component contribution to the overall observed emission is therefore small: consequently, since the ``back-side'' reprocessed component is expected to be a fraction of the front-side one, in this specific case it is basically negligible.

To further investigate this potential ``patchy torus'' scenario, in Section \ref{sec:variability} we divide our \NuSTAR\ and \XMM\ observations into three blocks and fit each of the sub-observations. As can be seen in Table \ref{Table:variability}, all the parameters are in agreement within the uncertainties, and no variability is therefore detected. This suggests that the timescale associated to a significant positional change of the clouds within the torus is much larger than $\sim$50\,ks or that the torus is more uniformly distributed than what the best-fit model suggests. However, the lack of significant variability in a 50\,ks observation provides us with a way to set a lower limit on the size of obscuring clouds. Assuming that the obscuring clouds are $r$ = 1\,pc away from the accreting SMBH \citep[see e.g.,][]{Naure_astro2017}. The mass of the super massive black hole in NGC 1358 is log(M$_{\rm BH}$/M${_\sun}$) = 7.88 \citep{Woo02}, such that the velocity of the clouds is v$_{\rm cloud}$ = (GM$_{\rm BH}/r$)$^{1/2}$ = 570\,km\,s$^{-1}$, which is in line with the FWHM velocity obtained from the velocity broadening of the emission lines. Therefore, the lower limit to the radius of the obscuring cloud should be R$_{\rm cloud}$ = v$_{\rm cloud}~\times$ 50\,ks = 3 $\times$ 10$^7$\,km $\sim$43\,R$_\sun$.

\subsection{Eddington ratio and mass accretion rate}
Finally, we analyzed our X-ray data to derive the Eddington ratio and mass accretion rate of the SMBH in NGC 1358. The Eddington ratio is a measurement of the SMBH accretion efficiency, and is defined as $\lambda_{\rm Edd}$ = $L_{\rm bol}/L_{\rm Edd}$, i.e., as the ratio between the bolometric luminosity, $L_{\rm bol}$, and the so-called Eddington luminosity, $L_{\rm Edd}$ = 4$\pi$GM$_{\rm BH}$m$_p$c/$\sigma_T$, where M$_{\rm BH}$ is the SMBH mass and m$_p$ is the mass of proton.

\citet{Woo02} report a measurement of both the black hole mass and the bolometric luminosity of NGC 1358. The mass of the black hole is obtained by the correlation between the black hole mass with stellar velocity dispersion, and the bolometric luminosity is obtained through direct integration of the spectral energy distribution. Based on these values, the Eddington ratio of NGC 1358 is $\lambda_{\rm Edd}$ = 2.5 $\times$ 10$^{-2}$. 

Our high-quality X-ray data allows us to recompute L$\rm _{bol}$, extrapolating it from our best-fit 2-10\,keV intrinsic luminosity, L$_{\rm int, 2-10\,keV}$ = 2.06$_{-0.11}^{+0.11}$ $\times$ 10$^{43}$\,erg\,s$^{-1}$. We use the bolometric correction of \citet[][Equation 21]{Lumicorrect}, finding a bolometric luminosity L$\rm _{bol}$ = 4.36$_{-0.30}^{+0.29}$ $\times$ 10$^{44}$\,erg\,s$^{-1}$. Thus the corresponding Eddington efficiency is $\lambda_{\rm Edd}$ $\sim$4.7$_{-0.3}^{+0.3}$ $\times$ 10$^{-2}$, which is typical for AGN in the range [0.001--1], using the black hole mass value from \citet{Woo02}.
This result is also consistent with \cite{Marinucci12}, which reported an Eddington ratio of $\lambda_{\rm Edd}$ = 1.62 $\times$ 10$^{-2}$, relying on an estimate of the black hole mass of log(M$_{\rm BH}$/M${_\sun}$) = 7.99, and a bolometric luminosity, L$\rm _{bol}$ = 2.04 $\times$ 10$^{44}$\,erg\,s$^{-1}$.

\citet{Schnorr17} estimated a NGC 1358 ionized mass inflow rate (excluding the neutral and molecular gas, being only the lower limit of the total mass inflow) of $\dot M_{\rm in}~\approx1.5\times10^{-2}$ M$_\sun$ yr$^{-1}$ in the inner 180\,pc. The authors also report the mass accretion rate, obtained using the [OIII] luminosity from \citet{gu02} and the bolometric correction from \citet{Lamastra09}, being $\dot m_{\rm acc}$ $\sim0.9\times10^{-4}$ M$_\sun$ yr$^{-1}$, which is 160 times smaller than the mass inflow rate of NGC 1358. According to the relationship between the mass accretion rate and the bolometric luminosity, $\dot m_{\rm acc}$ = L$_{\rm bol}$/$\eta$c$^2$, where $\eta$ is the efficiency that converts the rest mass energy of accreted material into radiation and is assumed to be $\eta$ = 0.1 \citep{Frank02}, we also estimate the mass accretion rate in NGC 1358, which is $\dot m_{\rm acc}$ = [7.2--8.2] $\times10^{-2}$ M$_\sun$ yr$^{-1}$ adopting the bolometric luminosity from our best-fit, which is in the same order with the mass inflow rate reported in \citet{Schnorr17}: the large difference between the mass accretion rate and the mass inflow rate reported in \citet{Schnorr17} could be diminished when the bolometric luminosity is measured from the 2--10\,keV X-ray band rather than from the [OIII] luminosity.

As shown here, \NuSTAR\ and \XMM\ are instrumental to identify and study CT-AGNs in details. We envision that extending these studies to most CT-AGN known in the Local Universe will allow us to shed light on the, so far elusive, population of CT-AGN.

%
%
\acknowledgments
X.Z. thanks the referee for their detailed and useful comments, which helped in significantly improving the paper. X.Z., S.M. and M.A. acknowledge NASA funding under contract 80NSSC17K0635. \NuSTAR\ is a project led by the California Institute of Technology (Caltech), managed by the Jet Propulsion Laboratory (JPL), and funded by the National Aeronautics and Space Administration (NASA). We thank the NuSTAR Operations, Software and Calibrations teams for support with these observations. This research has made use of the NuSTAR Data Analysis Software (NuSTARDAS) jointly developed by the ASI Science Data Center (ASDC, Italy) and the California Institute of Technology (USA). This research has made use of data and/or software provided by the High Energy Astrophysics Science Archive Research Center (HEASARC), which is a service of the Astrophysics Science Division at NASA/GSFC and the High Energy Astrophysics Division of the Smithsonian Astrophysical Observatory.

\bibliographystyle{aa}

\begin{thebibliography}{66}
\expandafter\ifx\csname natexlab\endcsname\relax\def\natexlab#1{#1}\fi

\bibitem[{Ajello {et~al.}(2008)Ajello, Greiner, Sato, Willis, Kanbach, Strong,
  Diehl, Hasinger, Gehrels, Markwardt, \& Tueller}]{Ajello08}
Ajello, M., Greiner, J., Sato, G., {et~al.} 2008, The Astrophysical Journal,
  689, 666

\bibitem[{Alexander {et~al.}(2003)Alexander, Bauer, Brandt, Schneider,
  Hornschemeier, Vignali, Barger, Broos, Cowie, Garmire, Townsley, Bautz,
  Chartas, \& Sargent}]{Alexander03}
Alexander, D.~M., Bauer, F.~E., Brandt, W.~N., {et~al.} 2003, The Astronomical
  Journal, 126, 539

\bibitem[{Almeida \& Ricci(2017)}]{Natureastro2017}
Almeida, C.~R. \& Ricci, C. 2017, Nature Astronomy, 1, 679

\bibitem[{Anders \& Grevesse(1989)}]{Anders1989}
Anders, E. \& Grevesse, N. 1989, Geochimica et Cosmochimica Acta, 53, 197

\bibitem[{Annuar {et~al.}(2017)Annuar, Alexander, Gandhi, Lansbury, Asmus,
  Ballantyne, Bauer, Boggs, Boorman, Brandt, Brightman, Christensen, Craig,
  Farrah, Goulding, Hailey, Harrison, Koss, LaMassa, Murray, Ricci, Rosario,
  Stanley, Stern, \& Zhang}]{Annuar17}
Annuar, A., Alexander, D.~M., Gandhi, P., {et~al.} 2017, The Astrophysical
  Journal, 836, 165

\bibitem[{Annuar {et~al.}(2015)Annuar, Gandhi, Alexander, Lansbury,
  Ar{\'e}valo, Ballantyne, Balokovi{\'c}, Bauer, Boggs, Brandt, Brightman,
  Christensen, Craig, Moro, Hailey, Harrison, Hickox, Matt, Puccetti, Ricci,
  Rigby, Stern, Walton, Zappacosta, \& Zhang}]{Annuar15}
Annuar, A., Gandhi, P., Alexander, D.~M., {et~al.} 2015, The Astrophysical
  Journal, 815, 36

\bibitem[{{Antonucci}(1993)}]{Antonucci1993}
{Antonucci}, R. 1993, Annu. Rev. Astron. Astrophys., 31, 473

\bibitem[{{Arnaud}(1996)}]{Arnaud1996}
{Arnaud}, K.~A. 1996, Astronomical Data Analysis Software and Systems V, 101,
  17

\bibitem[{Asmus {et~al.}(2015)Asmus, Gandhi, H{\"o}nig, Smette, \&
  Duschl}]{Asmus15}
Asmus, D., Gandhi, P., H{\"o}nig, S.~F., Smette, A., \& Duschl, W.~J. 2015,
  Monthly Notices of the Royal Astronomical Society, 454, 766

\bibitem[{Balokovi{\'c} {et~al.}(2018)Balokovi{\'c}, Brightman, Harrison,
  Comastri, Ricci, Buchner, Gandhi, Farrah, \& Stern}]{Borus}
Balokovi{\'c}, M., Brightman, M., Harrison, F.~A., {et~al.} 2018, The
  Astrophysical Journal, 854, 42

\bibitem[{Balokovi{\'c} {et~al.}(2014)Balokovi{\'c}, Comastri, Harrison,
  Alexander, Ballantyne, Bauer, Boggs, Brandt, Brightman, Christensen, Craig,
  Moro, Gandhi, Hailey, Koss, Lansbury, Luo, Madejski, Marinucci, Matt,
  Markwardt, Puccetti, Reynolds, Risaliti, Rivers, Stern, Walton, \&
  Zhang}]{Balokovic14}
Balokovi{\'c}, M., Comastri, A., Harrison, F.~A., {et~al.} 2014, The
  Astrophysical Journal, 794, 111

\bibitem[{Barthelmy {et~al.}(2005)Barthelmy, Barbier, Cummings, Fenimore,
  Gehrels, Hullinger, Krimm, Markwardt, Palmer, Parsons, Sato, Suzuki,
  Takahashi, Tashiro, \& Tueller}]{Barthelmy2005}
Barthelmy, S.~D., Barbier, L.~M., Cummings, J.~R., {et~al.} 2005, Space Science
  Reviews, 120, 143

\bibitem[{Bauer {et~al.}(2015)Bauer, Ar{\'e}valo, Walton, Koss, Puccetti,
  Gandhi, Stern, Alexander, Balokovi{\'c}, Boggs, Brandt, Brightman,
  Christensen, Comastri, Craig, Moro, Hailey, Harrison, Hickox, Luo, Markwardt,
  Marinucci, Matt, Rigby, Rivers, Saez, Treister, Urry, \& Zhang}]{Bauer15}
Bauer, F.~E., Ar{\'e}valo, P., Walton, D.~J., {et~al.} 2015, The Astrophysical
  Journal, 812, 116

\bibitem[{{Boller} {et~al.}(2016){Boller}, {Freyberg, M. J.}, {Tr\"umper, J.},
  {Haberl, F.}, {Voges, W.}, \& {Nandra, K.}}]{Boller16}
{Boller}, T., {Freyberg, M. J.}, {Tr\"umper, J.}, {et~al.} 2016, A\&A, 588,
  A103

\bibitem[{Brightman \& Nandra(2011)}]{BNtorus}
Brightman, M. \& Nandra, K. 2011, Monthly Notices of the Royal Astronomical
  Society, 413, 1206

\bibitem[{Buchner {et~al.}(2015)Buchner, Georgakakis, Nandra, Brightman,
  Menzel, Liu, Hsu, Salvato, Rangel, Aird, Merloni, \& Ross}]{Buchner2015}
Buchner, J., Georgakakis, A., Nandra, K., {et~al.} 2015, The Astrophysical
  Journal, 802, 89

\bibitem[{Burlon {et~al.}(2011)Burlon, Ajello, Greiner, Comastri, Merloni, \&
  Gehrels}]{Burlon11}
Burlon, D., Ajello, M., Greiner, J., {et~al.} 2011, The Astrophysical Journal,
  728, 58

\bibitem[{Elitzur \& Shlosman(2006)}]{Elitzur06}
Elitzur, M. \& Shlosman, I. 2006, The Astrophysical Journal Letters, 648, L101

\bibitem[{Elvis {et~al.}(1978)Elvis, Maccacaro, Wilson, Ward, Penston, Fosbury,
  \& Perola}]{Elvis1978}
Elvis, M., Maccacaro, T., Wilson, A.~S., {et~al.} 1978, Monthly Notices of the
  Royal Astronomical Society, 183, 129

\bibitem[{{Frank} {et~al.}(2002){Frank}, {King}, \& {Raine}}]{Frank02}
{Frank}, J., {King}, A., \& {Raine}, D.~J. 2002, Accretion Power in
  Astrophysics: Third Edition (Cambridge, UK: Cambridge University Press)

\bibitem[{Furui {et~al.}(2016)Furui, Fukazawa, Odaka, Kawaguchi, Ohno, \&
  Hayashi}]{Furui16}
Furui, S., Fukazawa, Y., Odaka, H., {et~al.} 2016, The Astrophysical Journal,
  818, 164

\bibitem[{Gandhi \& Fabian(2003)}]{Gandhi03}
Gandhi, P. \& Fabian, A.~C. 2003, Monthly Notices of the Royal Astronomical
  Society, 339, 1095

\bibitem[{{Georgantopoulos} \& {Akylas}(2010)}]{Georgantopoulos10}
{Georgantopoulos}, I. \& {Akylas}, A. 2010, A\&A, 509, A38

\bibitem[{{Georgantopoulos} {et~al.}(2013){Georgantopoulos}, {Comastri, A.},
  {Vignali, C.}, {Ranalli, P.}, {Rovilos, E.}, {Iwasawa, K.}, {Gilli, R.},
  {Cappelluti, N.}, {Carrera, F.}, {Fritz, J.}, {Brusa, M.}, {Elbaz, D.},
  {Mullaney, R. J.}, {Castello-Mor, N.}, {Barcons, X.}, {Tozzi, P.}, {Balestra,
  I.}, \& {Falocco, S.}}]{Georgantopoulos2013}
{Georgantopoulos}, I., {Comastri, A.}, {Vignali, C.}, {et~al.} 2013, A\&A, 555,
  A43

\bibitem[{{Gilli} {et~al.}(2007){Gilli}, {Comastri, A.}, \& {Hasinger,
  G.}}]{gilli07}
{Gilli}, R., {Comastri, A.}, \& {Hasinger, G.} 2007, A\&A, 463, 79

\bibitem[{Gu \& Huang(2002)}]{gu02}
Gu, Q. \& Huang, J. 2002, The Astrophysical Journal, 579, 205

\bibitem[{Harrison {et~al.}(2013)Harrison, Craig, Christensen, Hailey, Zhang,
  Boggs, Stern, Cook, Forster, Giommi, Grefenstette, Kim, Kitaguchi, Koglin,
  Madsen, Mao, Miyasaka, Mori, Perri, Pivovaroff, Puccetti, Rana, Westergaard,
  Willis, Zoglauer, An, Bachetti, Barri{\`e}re, Bellm, Bhalerao, Brejnholt,
  Fuerst, Liebe, Markwardt, Nynka, Vogel, Walton, Wik, Alexander, Cominsky,
  Hornschemeier, Hornstrup, Kaspi, Madejski, Matt, Molendi, Smith, Tomsick,
  Ajello, Ballantyne, Balokovi{\'c}, Barret, Bauer, Blandford, Brandt,
  Brenneman, Chiang, Chakrabarty, Chenevez, Comastri, Dufour, Elvis, Fabian,
  Farrah, Fryer, Gotthelf, Grindlay, Helfand, Krivonos, Meier, Miller,
  Natalucci, Ogle, Ofek, Ptak, Reynolds, Rigby, Tagliaferri, Thorsett,
  Treister, \& Urry}]{harrison}
Harrison, F.~A., Craig, W.~W., Christensen, F.~E., {et~al.} 2013, The
  Astrophysical Journal, 770, 103

\bibitem[{Heckman {et~al.}(2005)Heckman, Ptak, Hornschemeier, \&
  Kauffmann}]{Heckman05}
Heckman, T.~M., Ptak, A., Hornschemeier, A., \& Kauffmann, G. 2005, The
  Astrophysical Journal, 634, 161

\bibitem[{Ikeda {et~al.}(2009)Ikeda, Awaki, \& Terashima}]{Shinya09}
Ikeda, S., Awaki, H., \& Terashima, Y. 2009, The Astrophysical Journal, 692,
  608

\bibitem[{Jaffe {et~al.}(2004)Jaffe, Meisenheimer, R{\"o}ttgering, Leinert,
  Richichi, Chesneau, Fraix-Burnet, Glazenborg-Kluttig, Granato, Graser,
  Heijligers, K{\"o}hler, Malbet, Miley, Paresce, Pel, Perrin, Przygodda,
  Schoeller, Sol, Waters, Weigelt, Woillez, \& de~Zeeuw}]{Jaffe2004}
Jaffe, W., Meisenheimer, K., R{\"o}ttgering, H. J.~A., {et~al.} 2004, Nature,
  429, 47 EP

\bibitem[{{Jansen} {et~al.}(2001){Jansen}, {Lumb, D.}, {Altieri, B.}, {Clavel,
  J.}, {Ehle, M.}, {Erd, C.}, {Gabriel, C.}, {Guainazzi, M.}, {Gondoin, P.},
  {Much, R.}, {Munoz, R.}, {Santos, M.}, {Schartel, N.}, {Texier, D.}, \&
  {Vacanti, G.}}]{SAS}
{Jansen}, F., {Lumb, D.}, {Altieri, B.}, {et~al.} 2001, A\&A, 365, L1

\bibitem[{{Kalberla} {et~al.}(2005){Kalberla}, {Burton, W. B.}, {Hartmann,
  Dap}, {Arnal, E. M.}, {Bajaja, E.}, {Morras, R.}, \& {P{\"o}ppel, W. G.
  L.}}]{Kalberla05}
{Kalberla}, P. M.~W., {Burton, W. B.}, {Hartmann, Dap}, {et~al.} 2005, A\&A,
  440, 775

\bibitem[{Koss {et~al.}(2016)Koss, Assef, Balokovi{\'c}, Stern, Gandhi,
  Lamperti, Alexander, Ballantyne, Bauer, Berney, Brandt, Comastri, Gehrels,
  Harrison, Lansbury, Markwardt, Ricci, Rivers, Schawinski, Trakhtenbrot,
  Treister, \& Urry}]{koss2016}
Koss, M.~J., Assef, R., Balokovi{\'c}, M., {et~al.} 2016, The Astrophysical
  Journal, 825, 85

\bibitem[{{Krolik} \& {Begelman}(1988)}]{Krolik1988}
{Krolik}, J.~H. \& {Begelman}, M.~C. 1988, Astrophysical Journal, 329, 702

\bibitem[{{Lamastra} {et~al.}(2009){Lamastra}, {Bianchi, S.}, {Matt, G.},
  {Perola, G. C.}, {Barcons, X.}, \& {Carrera, F. J.}}]{Lamastra09}
{Lamastra}, A., {Bianchi, S.}, {Matt, G.}, {et~al.} 2009, A\&A, 504, 73

\bibitem[{{Lanzuisi} {et~al.}(2015){Lanzuisi}, {Ranalli, P.}, {Georgantopoulos,
  I.}, {Georgakakis, A.}, {Delvecchio, I.}, {Akylas, T.}, {Berta, S.},
  {Bongiorno, A.}, {Brusa, M.}, {Cappelluti, N.}, {Civano, F.}, {Comastri, A.},
  {Gilli, R.}, {Gruppioni, C.}, {Hasinger, G.}, {Iwasawa, K.}, {Koekemoer, A.},
  {Lusso, E.}, {Marchesi, S.}, {Mainieri, V.}, {Merloni, A.}, {Mignoli, M.},
  {Piconcelli, E.}, {Pozzi, F.}, {Rosario, D. J.}, {Salvato, M.}, {Silverman,
  J.}, {Trakhtenbrot, B.}, {Vignali, C.}, \& {Zamorani, G.}}]{Lanzuisi2105}
{Lanzuisi}, G., {Ranalli, P.}, {Georgantopoulos, I.}, {et~al.} 2015, A\&A, 573,
  A137

\bibitem[{Liu \& Li(2014)}]{Liu14}
Liu, Y. \& Li, X. 2014, The Astrophysical Journal, 787, 52

\bibitem[{Magdziarz \& Zdziarski(1995)}]{pexrav}
Magdziarz, P. \& Zdziarski, A.~A. 1995, Monthly Notices of the Royal
  Astronomical Society, 273, 837

\bibitem[{Marchesi {et~al.}(2017{\natexlab{a}})Marchesi, Ajello, Comastri,
  Cusumano, Parola, \& Segreto}]{marchesi2017APJ}
Marchesi, S., Ajello, M., Comastri, A., {et~al.} 2017{\natexlab{a}}, The
  Astrophysical Journal, 836, 116

\bibitem[{Marchesi {et~al.}(2018)Marchesi, Ajello, Marcotulli, Comastri,
  Lanzuisi, \& Vignali}]{Marchesi2018}
Marchesi, S., Ajello, M., Marcotulli, L., {et~al.} 2018, The Astrophysical
  Journal, 854, 49

\bibitem[{Marchesi {et~al.}(2016)Marchesi, Lanzuisi, Civano, Iwasawa, Suh,
  Comastri, Zamorani, Allevato, Griffiths, Miyaji, Ranalli, Salvato,
  Schawinski, Silverman, Treister, Urry, \& Vignali}]{marchesi2016}
Marchesi, S., Lanzuisi, G., Civano, F., {et~al.} 2016, The Astrophysical
  Journal, 830, 100

\bibitem[{Marchesi {et~al.}(2017{\natexlab{b}})Marchesi, Tremblay, Ajello,
  Marcotulli, Paggi, Cusumano, Parola, \& Segreto}]{Stefano2017}
Marchesi, S., Tremblay, L., Ajello, M., {et~al.} 2017{\natexlab{b}}, The
  Astrophysical Journal, 848, 53

\bibitem[{Marconi {et~al.}(2004)Marconi, Risaliti, Gilli, Hunt, Maiolino, \&
  Salvati}]{Lumicorrect}
Marconi, A., Risaliti, G., Gilli, R., {et~al.} 2004, Monthly Notices of the
  Royal Astronomical Society, 351, 169

\bibitem[{Marinucci {et~al.}(2012)Marinucci, Bianchi, Nicastro, Matt, \&
  Goulding}]{Marinucci12}
Marinucci, A., Bianchi, S., Nicastro, F., Matt, G., \& Goulding, A.~D. 2012,
  The Astrophysical Journal, 748, 130

\bibitem[{Matt \& Fabian(1994)}]{Matt1994}
Matt, G. \& Fabian, A.~C. 1994, Monthly Notices of the Royal Astronomical
  Society, 267, 187

\bibitem[{Mendoza-Castrej{\'o}n {et~al.}(2015)Mendoza-Castrej{\'o}n, Dultzin,
  Krongold, Gonz{\'a}lez, \& Elitzur}]{Mendoza15}
Mendoza-Castrej{\'o}n, S., Dultzin, D., Krongold, Y., Gonz{\'a}lez, J.~J., \&
  Elitzur, M. 2015, Monthly Notices of the Royal Astronomical Society, 447,
  2437

\bibitem[{{Mewe} {et~al.}(1985){Mewe}, {Gronenschild}, \& {van den
  Oord}}]{mekal}
{Mewe}, R., {Gronenschild}, E.~H.~B.~M., \& {van den Oord}, G.~H.~J. 1985,
  A\&AS, 62, 197

\bibitem[{Murphy \& Yaqoob(2009)}]{MYTorus2009}
Murphy, K.~D. \& Yaqoob, T. 2009, Monthly Notices of the Royal Astronomical
  Society, 397, 1549

\bibitem[{Nenkova {et~al.}(2008)Nenkova, Sirocky, Ivezi{\'c}, \&
  Elitzur}]{Nenkova08}
Nenkova, M., Sirocky, M.~M., Ivezi{\'c}, {\v Z}., \& Elitzur, M. 2008, The
  Astrophysical Journal, 685, 147

\bibitem[{Puccetti {et~al.}(2014)Puccetti, Comastri, Fiore, Ar{\'e}valo,
  Risaliti, Bauer, Brandt, Stern, Harrison, Alexander, Boggs, Christensen,
  Craig, Gandhi, Hailey, Koss, Lansbury, Luo, Madejski, Matt, Walton, \&
  Zhang}]{puccetti14}
Puccetti, S., Comastri, A., Fiore, F., {et~al.} 2014, The Astrophysical
  Journal, 793, 26

\bibitem[{Ricci {et~al.}(2015)Ricci, Ueda, Koss, Trakhtenbrot, Bauer, \&
  Gandhi}]{Ricci15}
Ricci, C., Ueda, Y., Koss, M.~J., {et~al.} 2015, The Astrophysical Journal
  Letters, 815, L13

\bibitem[{{Ricci} {et~al.}(2011){Ricci}, {Walter, R.}, {Courvoisier, T. J.-L.},
  \& {Paltani, S.}}]{Ricci11}
{Ricci}, C., {Walter, R.}, {Courvoisier, T. J.-L.}, \& {Paltani, S.} 2011,
  A\&A, 532, A102

\bibitem[{Risaliti {et~al.}(1999)Risaliti, Maiolino, \& Salvati}]{risaliti1999}
Risaliti, G., Maiolino, R., \& Salvati, M. 1999, The Astrophysical Journal,
  522, 157

\bibitem[{Schnorr-M{\"u}ller {et~al.}(2017)Schnorr-M{\"u}ller,
  Storchi-Bergmann, Nagar, Robinson, \& Lena}]{Schnorr17}
Schnorr-M{\"u}ller, A., Storchi-Bergmann, T., Nagar, N.~M., Robinson, A., \&
  Lena, D. 2017, Monthly Notices of the Royal Astronomical Society, 471, 3888

\bibitem[{Stalevski {et~al.}(2016)Stalevski, Ricci, Ueda, Lira, Fritz, \&
  Baes}]{Stalevski16}
Stalevski, M., Ricci, C., Ueda, Y., {et~al.} 2016, Monthly Notices of the Royal
  Astronomical Society, 458, 2288

\bibitem[{{Theureau} {et~al.}(1998){Theureau}, {Bottinelli, L.},
  {Coudreau-Durand, N.}, {Gouguenheim, L.}, {Hallet, N.}, {Loulergue, M.},
  {Paturel, G.}, \& {Teerikorpi, P.}}]{Theureau1998}
{Theureau}, G., {Bottinelli, L.}, {Coudreau-Durand, N.}, {et~al.} 1998, Astron.
  Astrophys. Suppl. Ser., 130, 333

\bibitem[{Treister {et~al.}(2009)Treister, Urry, \& Virani}]{Treister09}
Treister, E., Urry, C.~M., \& Virani, S. 2009, The Astrophysical Journal, 696,
  110

\bibitem[{Ueda {et~al.}(2014)Ueda, Akiyama, Hasinger, Miyaji, \&
  Watson}]{Ueda14}
Ueda, Y., Akiyama, M., Hasinger, G., Miyaji, T., \& Watson, M.~G. 2014, The
  Astrophysical Journal, 786, 104

\bibitem[{Ursini {et~al.}(2018)Ursini, Bassani, Panessa, Bazzano, Bird,
  Malizia, \& Ubertini}]{Ursini18}
Ursini, F., Bassani, L., Panessa, F., {et~al.} 2018, Monthly Notices of the
  Royal Astronomical Society, 474, 5684

\bibitem[{Verner {et~al.}(1996)Verner, Ferland, Korista, \&
  Yakovlev}]{Verner1996}
Verner, D., Ferland, G., Korista, K., \& Yakovlev, D. 1996, Astrophysical
  Journal, 465, 487

\bibitem[{{Whittle}(1992)}]{Whittle1992}
{Whittle}, M. 1992, \apjs, 79, 49

\bibitem[{{Winkler} {et~al.}(2003){Winkler}, {T. J.-L. Courvoisier}, {Di Cocco,
  G.}, {Gehrels, N.}, {Gim\'enez, A.}, {Grebenev, S.}, {Hermsen, W.},
  {Mas-Hesse, J. M.}, {Lebrun, F.}, {Lund, N.}, {Palumbo, G. G. C.}, {Paul,
  J.}, {Roques, J.-P.}, {Schnopper, H.}, {Sch\"onfelder, V.}, {Sunyaev, R.},
  {Teegarden, B.}, {Ubertini, P.}, {Vedrenne, G.}, \& {Dean, A.
  J.}}]{Winkler2003}
{Winkler}, C., {T. J.-L. Courvoisier}, {Di Cocco, G.}, {et~al.} 2003, A\&A,
  411, L1

\bibitem[{Woo \& Urry(2002)}]{Woo02}
Woo, J.-H. \& Urry, C.~M. 2002, The Astrophysical Journal, 579, 530

\bibitem[{Wright {et~al.}(2010)Wright, Eisenhardt, Mainzer, Ressler, Cutri,
  Jarrett, Kirkpatrick, Padgett, McMillan, Skrutskie, Stanford, Cohen, Walker,
  Mather, Leisawitz, III, McLean, Benford, Lonsdale, Blain, Mendez, Irace,
  Duval, Liu, Royer, Heinrichsen, Howard, Shannon, Kendall, Walsh, Larsen,
  Cardon, Schick, Schwalm, Abid, Fabinsky, Naes, \& Tsai}]{wright10}
Wright, E.~L., Eisenhardt, P. R.~M., Mainzer, A.~K., {et~al.} 2010, The
  Astronomical Journal, 140, 1868

\bibitem[{Yaqoob(2012)}]{MYTorus2012}
Yaqoob, T. 2012, Monthly Notices of the Royal Astronomical Society, 423, 3360

\bibitem[{Yaqoob {et~al.}(2015)Yaqoob, Tatum, Scholtes, Gottlieb, \&
  Turner}]{MYTorus2015}
Yaqoob, T., Tatum, M.~M., Scholtes, A., Gottlieb, A., \& Turner, T.~J. 2015,
  Monthly Notices of the Royal Astronomical Society, 454, 973

\end{thebibliography}

\end{document}